\documentclass[]{aa}

\usepackage{epsfig}    
\usepackage{lscape}
\usepackage{natbib}
\bibpunct{(}{)}{;}{a}{}{,}    
\usepackage{graphicx,graphics}
\usepackage{color} 
\usepackage{gensymb}
\usepackage{times}
\usepackage{amsmath}
\usepackage{amssymb}
\usepackage[normalem]{ulem}
\begin{document}

\title{Characterizing the radial oxygen abundance distribution in disk galaxies}

\author{
        I.~A.~Zinchenko\inst{\ref{MAO},\ref{ARI}} \and
        A.~Just\inst{\ref{ARI}}  \and     
        L.~S.~Pilyugin\inst{\ref{MAO},\ref{ARI}} \and
        M.~A.~Lara-Lopez\inst{\ref{DARK}}
       }
       
\institute{
Main Astronomical Observatory, National Academy of Sciences of Ukraine, 
27 Akademika Zabolotnoho St., 03680, Kyiv, Ukraine \label{MAO}
\and
Astronomisches Rechen-Institut, Zentrum f\"{u}r Astronomie 
der Universit\"{a}t Heidelberg, 
M\"{o}nchhofstr.\ 12--14, 69120 Heidelberg, Germany \label{ARI} 
\and
Dark Cosmology Centre, Niels Bohr Institute, University of Copenhagen, 
Juliane Maries Vej 30, DK-2100 Copenhagen, Denmark \label{DARK}
}

\abstract{The relation between the radial oxygen abundance distribution (gradient) and other parameters of a galaxy 
such as mass, Hubble type, and a bar strength, remains unclear although a large amount of observational
data have been obtained in the past years.}
{We examine the possible dependence of the radial oxygen abundance distribution on non-axisymmetrical structures (bar/spirals) 
and other macroscopic parameters such as
the mass, the optical radius $R_{25}$, the color $g-r$, and the surface brightness of the galaxy. 
A sample of disk galaxies from the third data release of the Calar Alto Legacy Integral Field Area Survey (CALIFA DR3) is considered.}
{We adopted the Fourier amplitude A$_2$ of the surface brightness as a quantitative characteristic 
of the strength of non-axisymmetric structures in a galactic disk, in addition to the commonly used morphologic division
for A, AB, and B types based on the Hubble classification. To distinguish changes in local oxygen abundance caused by 
the non-axisymmetrical structures, the multiparametric 
mass--metallicity relation
was constructed as a function of parameters such as the bar/spiral pattern strength, the disk size, 
color index $g-r$ in the Sloan Digital Sky Survey (SDSS) bands, and central surface brightness of the disk. 
The gas-phase oxygen abundance gradient is determined by using the R calibration.
}
{We find that there is no significant impact of the non-axisymmetric structures such as a bar 
and/or spiral patterns on the local oxygen abundance and radial oxygen abundance gradient of 
disk galaxies. Galaxies with higher mass, however, exhibit flatter oxygen abundance gradients 
in units of dex/kpc, but this effect is significantly less prominent for the oxygen abundance gradients in units of dex/$R_{25}$
and almost disappears when the inner parts are avoided ($R > 0.25R_{25}$). 
We show that the oxygen abundance in the central part of the galaxy depends  
neither on the optical radius $R_{25}$ nor on the color $g-r$ or the surface brightness of the galaxy.
Instead, outside the central part of the galaxy, the oxygen abundance increases with $g-r$ value 
and central surface brightness of the disk. 
}
{}



\titlerunning{Oxygen abundance gradient in CALIFA galaxies}
\authorrunning{Zinchenko et al.}
\maketitle

\section{Introduction}

It is known that the chemical abundances of heavy elements decrease with increasing galactocentric 
distance in the disks of our Galaxy and other nearby spiral galaxies \citep[e.g.,][]{Mayor1976, 
Searle1971, Peimbert1979, Shields1978, Belfiore2017}. 
This trend is usually referred to as the radial metallicity gradient (RMG).
The oxygen abundance gradient in nearby galaxies, traced by H II regions,
ranges from $\sim$ 0 dex kpc$^{-1}$ to $-0.1$ dex kpc$^{-1}$
\citep[e.g.,][]{Vila-Costas1992, Zaritsky1994, Sanchez2013, Pilyugin2014a}.
However, the origin of the gradient and its relation to
other macroscopic parameters of galaxies are still open for discussion.

During the past decade, significant progress in the understanding of the main tendencies of 
chemical abundance distribution in galaxies became possible through integral field spectroscopy 
(IFS) data that were obtained in frameworks of large galaxy surveys. One such survey is the Calar Alto 
Legacy Integral Field Area Survey \citep[CALIFA;][]{Sanchez2012}. Using oxygen abundance measurements in a large number of H II regions 
in more than 300 spiral galaxies from the CALIFA survey, \citet{Sanchez2014} showed that disk galaxies in 
the local Universe exhibit a common characteristic gradient in the oxygen abundance up to two 
effective disk radii, regardless of the galaxy parameters.
The distribution of the abundance gradients is Gaussian around some mean value. Therefore, \citet{Sanchez2014}
suggested that the scatter around the mean value could be a result of random fluctuations.

\citet{Sanchez2014} claimed that this result contradicts several previous observational studies that concluded
that the slope of the gas-phase abundance gradient is related to other 
galactic properties, such as

\begin{itemize}
\item[--] morphology: the slopes of early-type spirals are more shallow than those of
          late-type galaxies (e.g., \citet{McCall1985};
          \citet{Vila-Costas1992});
\item[--] mass: the slopes of more massive spirals are more shallow than those of less
          massive galaxies (e.g., \citet{Zaritsky1994};
          \citet{Martin1994}; \citet{Garnett1998});
\item[--] presence of a bar: the slopes of barred galaxies are more shallow
          than than those of unbarred galaxies (e.g., \citet{Vila-Costas1992};
          \citet{Zaritsky1994}; \citet{Roy1996});
\item[--] interaction stage of the galaxies: evolved mergers present
          shallower slopes (e.g., \citet{Rich2012}).
\end{itemize}

In contrast, many authors found no correlation between abundance gradient 
(expressed in dex/$R_{25}$) and morphology \citep{Zaritsky1994,Pilyugin2014a,Sanchez2014,SanchezMenguiano2016}
and stellar mass \citep{Ho2015,SanchezMenguiano2016,SanchezMenguiano2018}. 
However, for the metallicity gradients expressed in dex/kpc, 
\citet{Ho2015} found that galaxies with lower mass tend to have 
steeper metallicity gradients on average.

Using the CALIFA survey, \citet{Sanchez-Blazquez2014} found no difference in 
the slope of the stellar-phase metallicity gradient for barred and unbarred galaxies. 
The same conclusion was made by \citet{Cheung2015}, who analyzed stacked spectra from the Sloan Digital Sky Survey (SDSS).

Based on a sample of 550 nearby galaxies with integral
field unit (IFU) data from the SDSS IV Mapping Nearby Galaxies at Apache Point Observatory survey \citep[SDSS/MaNGA;][]{Bundy2015,Blanton2017},
it has recently been found that the radial metallicity gradient steepens when the stellar mass increases 
up to $\log(M/M_{\odot}) = 10.5,$ but flattens for galaxies with higher masses \citep{Belfiore2017}.
However, using the same data from the SDSS-IV MaNGA survey, \citet{Lian2018} found that 
the stellar metallicity gradients tend to be mass dependent with steeper gradients in more massive galaxies, 
but they found no clear mass dependence for the gas metallicity gradient.
Moreover, a recent analysis of IFU data suggests that galaxies can present a different behavior of 
the radial abundance gradient in the inner and outer parts \citep{Sanchez2014,SanchezMenguiano2016,Belfiore2017,SanchezMenguiano2018}.

According to the current observational picture presented above, despite the wide range of metallicity gradients 
exhibited by galaxies, only a few studies point out the connection between radial metallicity gradient and galaxy mass. Thus, there is no solid confirmation of the 
correlation between oxygen abundance gradient and galaxy parameters (with the exception of mergers).

On the other hand, theoretical simulations based on pure N-body galactic disk dynamics and self-consistent 
chemodynamical models produce an appreciable influence of non-axisymmetric structures
(bar and spiral patterns) in the galactic disk, and interactions between galaxies.
Several theoretical studies have suggested various mechanisms of gas and star mixing 
in barred galaxies \citep{Athanassoula1992,Sellwood2002,Schonrich2009,Minchev2010,DiMatteo2013,Minchev2013,Minchev2014} that might lead to the flattening of the oxygen abundance distribution.
This scenario is consistent with observational results \citep{RuizLara2017} based on 
the analysis of the stellar content in a sample of CALIFA galaxies.
Thus, the redistribution of heavy elements produced by the radial migration of stars
and/or interstellar gas along the galactic disk is assumed to be the cause of a flattening of the metallicity gradient for the intermediate-age
stellar populations and interstellar gas.
However, \citet{Grand2016} demonstrated that corotating spiral arms do not change the radial metallicity gradient.

 \citet{Zinchenko2016} analyzed the azimuthal variation of the oxygen abundance in CALIFA galaxies 
and found no significant asymmetry in the azimuthal abundance distribution.
In this paper, we investigate the possible dependence of the oxygen abundance gradient in  
disk galaxies on a bar/spiral and other parameters of the galaxy, such as 
mass, size, color index, and surface brightness. We also study the mass--metallicity relation 
for the local oxygen abundance in the center and in intermediate and outer parts of the galaxy
as a multivariable relation including size, color index, surface brightness, and the bar/spiral 
pattern strength.

\section{Data}

\subsection{Sample}

We used publicly available spectra from the integral field spectroscopic CALIFA
survey data release 3 \citep[DR3;][]{Sanchez2016,Sanchez2012,CALIFA2014} based on
observations with the Potsdam Multi-Aperture Spectrophotometer (PMAS) and a specialized fiber-bundle, called PPAK (Pmas fiber PAcK), 
mounted on the Calar Alto 3.5-meter telescope. CALIFA DR3 provides wide-field IFU data for 667 objects in total.  
The data for each galaxy are presented by two spectral datacubes that cover the spectral regions of
4300--7000~\AA\ at a spectral resolution of $R \sim 850$ (setup V500) and of 3700--5000~\AA\ 
at $R \sim 1650$ (setup V1200). Currently, there are 446 COMB datacubes that are a combination of V500 and V1200 datacubes and 
cover the spectral range of 3700--7000~\AA\/. In this study we used COMB (a combination of V500+V1200)
datacubes.

The galaxy inclination $i$ and position angle of the major axis $PA$ were estimated from the 
analysis of the surface brightness image in the $r$ band of the SDSS. 
We applied the GALFIT code \citep{Peng2002,Peng2010} to simultaenously fit the image of each galaxy with 
bulge+disk (Sersic+exponential profile) profiles.
The position angle of the major axis and inclination of the exponential profile were adopted 
as the $PA$ and $i$ for all galaxies, with the exception of NGC~1070. 
Because the fit of the exponential profile for NGC~1070 is unreliable, the $PA$ and $i$ of the Sersic profile 
were adopted instead.

For further consideration, we selected galaxies with inclinations lower than $60\degree$,
which corresponds to a ratio of minor to major axis greater than $\sim0.5$.
The optical isophotal radius $R_{25}$ of a galaxy was determined from the analysis of the surface 
brightness profiles in the SDSS $g$ and $r$ bands converted into the Vega $B$ band. 
We used the stellar masses derived from UV-to-near-infrared (NIR) photometry, morphological type, bar, and merger classification
from the parameter tables described in \citet{CALIFA2014}. 
We adopted the distances from the NED database with flow corrections
for Virgo, the Great Attractor, and the Shapley Supercluster infall.
The adopted distances, morphological types, $PA$, $i$, stellar masses, $R_{25}$, and $g-r$ colors
are shown in Table~\ref{table:sample}.

\begin{figure}
\includegraphics[width=1.0\linewidth]{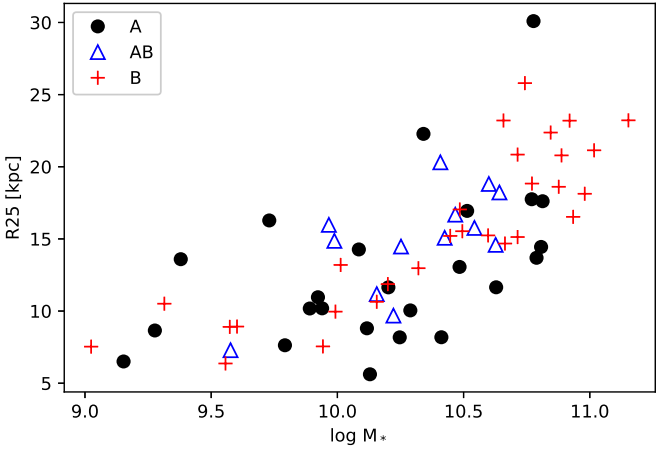}
\caption{Comparison of stellar mass and optical radius $R_{25}$ for our sample of galaxies.
Unbarred (A), barred (B), and intermediate (AB) galaxies are shown with different colors and symbols.
}
\label{figure:mass-R25}
\end{figure}

We selected only galaxies with no merger features, that is, galaxies that are classified as isolated. 
Galaxies with insufficient numbers of spaxels with measured oxygen abundance were excluded from our sample.
Thus, our final sample contains 66 galaxies with a radial oxygen abundance gradient derived by 
at least 10 spaxels that are distributed in a significantly wide radial range.
Figure~\ref{figure:mass-R25} shows the correlation between the stellar mass and the optical isophotal radius
for our sample of galaxies.

\begin{figure}
  \includegraphics[width=1.0\linewidth]{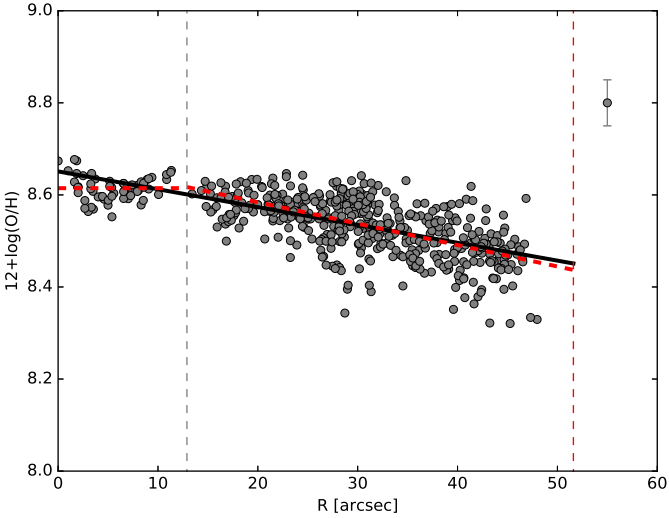}
\caption{Oxygen abundance gradient for NGC~257. Circles represent the oxygen abundance 
in the individual spaxels. Vertical dashed lines represent 0.25 and 1.0 fraction 
of optical radius $R_{25}$. The black solid line shows the best fit to all the data.
Dashed red lines correspond to the best fit of the oxygen abundance gradient in the 
inner ($R < 0.25 R_{25}$) and outer ($0.25 R_{25} < R < R_{25}$) parts of the galaxy. 
The typical uncertainty of the oxygen abundance determination for the single spaxel  
is presented in the top right corner.
}
\label{figure:NGC0257-grad}
\end{figure}

\subsection{Emission line fluxes and abundance determination}

The spectrum of each spaxel from the CALIFA DR3 datacubes was processed in the same way as 
described in \citet{Zinchenko2016}. Briefly, 
the stellar background in all spaxels was fit using the public version of the STARLIGHT code 
\citep{CidFernandes2005,Mateus2006,Asari2007} adapted for execution in the NorduGrid 
ARC\footnote{http://www.nordugrid.org/} environment of the Ukrainian National Grid.
We used a set of 45 synthetic simple stellar population (SSP) spectra with metallicities $Z = 0.004$, 0.02, and 0.05, 
and 15 ages from 1~Myr up to 13~Gyr  
from the evolutionary synthesis models of \citet{BC03} and the reddening law of \citet[]{CCM} with $R_V = 3.1$.
The resulting stellar radiation contribution was subtracted from the measured spectrum 
in order to determine the nebular emission spectrum.
For each spectrum, we measured the fluxes of the 
[O\,{\sc ii}]$\lambda$3727+$\lambda$3729, 
H$\beta$,  
[O\,{\sc iii}]$\lambda$4959, 
[O\,{\sc iii}]$\lambda$5007,
[N\,{\sc ii}]$\lambda$6548,
H$\alpha$,  
[N\,{\sc ii}]$\lambda$6584, 
[S\,{\sc ii}]$\lambda$6717, and 
[S\,{\sc ii}]$\lambda$6731 lines
using our code ELF3D for emission line fitting. Each line was fit with a Gaussian profile.
The measured line fluxes were corrected for interstellar reddening using the theoretical H$\alpha$ to H$\beta$ ratio  
(i.e., the standard  value of H$\alpha$/H$\beta$ = 2.86) and the analytical approximation of the Whitford interstellar 
reddening law from \citet{Izotov1994}.  When the measured value of H$\alpha$/H$\beta$ was lower than 2.86, the 
reddening was set to zero.

The [O\,{\sc iii}]$\lambda$5007 and $\lambda$4959 lines originate from transitions from the 
same energy level, therefore their flux ratio is determined only by the transition probability ratio, 
which is  very close to 3 \citep{Storey2000}. The stronger line [O\,{\sc iii}]$\lambda$5007 
is usually measured with higher precision than the weaker line [O\,{\sc iii}]$\lambda$4959. 
Therefore, the value of $R_3$ is estimated as $1.33$~[O\,{\sc iii}]$\lambda$5007, but not 
as a sum of the line fluxes.  
Similarly, the [N\,{\sc ii}]$\lambda$6584 and $\lambda$6548 lines also originate from transitions from the 
same energy level, and the transition probability ratio for these lines is again
close to 3 \citep{Storey2000}. The value of $N_2$ is therefore estimated as 
$1.33$~[N\,{\sc ii}]$\lambda$6584. Thus, the lines 
[O\,{\sc ii}]$\lambda$$\lambda$3727,3729,
H$\beta$,  
[O\,{\sc iii}]$\lambda$5007,
H$\alpha$,  
[N\,{\sc ii}]$\lambda$6584, 
[S\,{\sc ii}]$\lambda$6717, and
[S\,{\sc ii}]$\lambda$6731 
are used for the dereddening and the abundance determinations. 
The precision of the line flux is 
specified by the ratio of the flux to the flux error (parameter $\epsilon$). 
We selected spectra where the parameter $\epsilon \geq 4$ for each of these lines. 
To select spaxels associated with star-forming regions, we applied BPT diagram 
$\log$([O\,{\sc iii}]$\lambda$5007/H$\beta$) -- $\log$([N\,{\sc ii}]$\lambda\lambda$6584/H$\alpha$) \citep{BPT}.
To separate objects whose main ionization source are massive stars from those 
whose main ionization source are shocks of gas and/or active galactic nuclei (AGNs), we 
applied the criterion proposed by \cite{Kauffmann2003}.

The CALIFA spectra, as well as the spectra of other surveys that use  the IFU technique,
can be contaminated by diffuse ionized gas \citep[DIG;][]{Haffner2009,Belfiore2015,Lacerda2018}.
This can lead to an increasing emission in low-ionization lines and therefore to biases in the chemical 
abundances derived by strong-line methods. However, \citet{Pilyugin2018} showed that the 
dividing line proposed by \cite{Kauffmann2003} can be applied to reject spectra
with significant contamination by DIG.

To determine oxygen abundances, we used the new R calibration 
developed by \citet{PilyuginGrebel2016}.
In comparison with previous calibrations, this calibration is applicable for the whole range of
H~II region metallicities and provides high-precision oxygen abundances  
with measurement errors smaller than 0.1~dex and a metallicity scale bounded to the direct T$_e$ method.
Thus, the oxygen abundance was calculated for each spaxel using the R calibration. 

\begin{figure*}
\begin{center}
\includegraphics[width=0.80\linewidth]{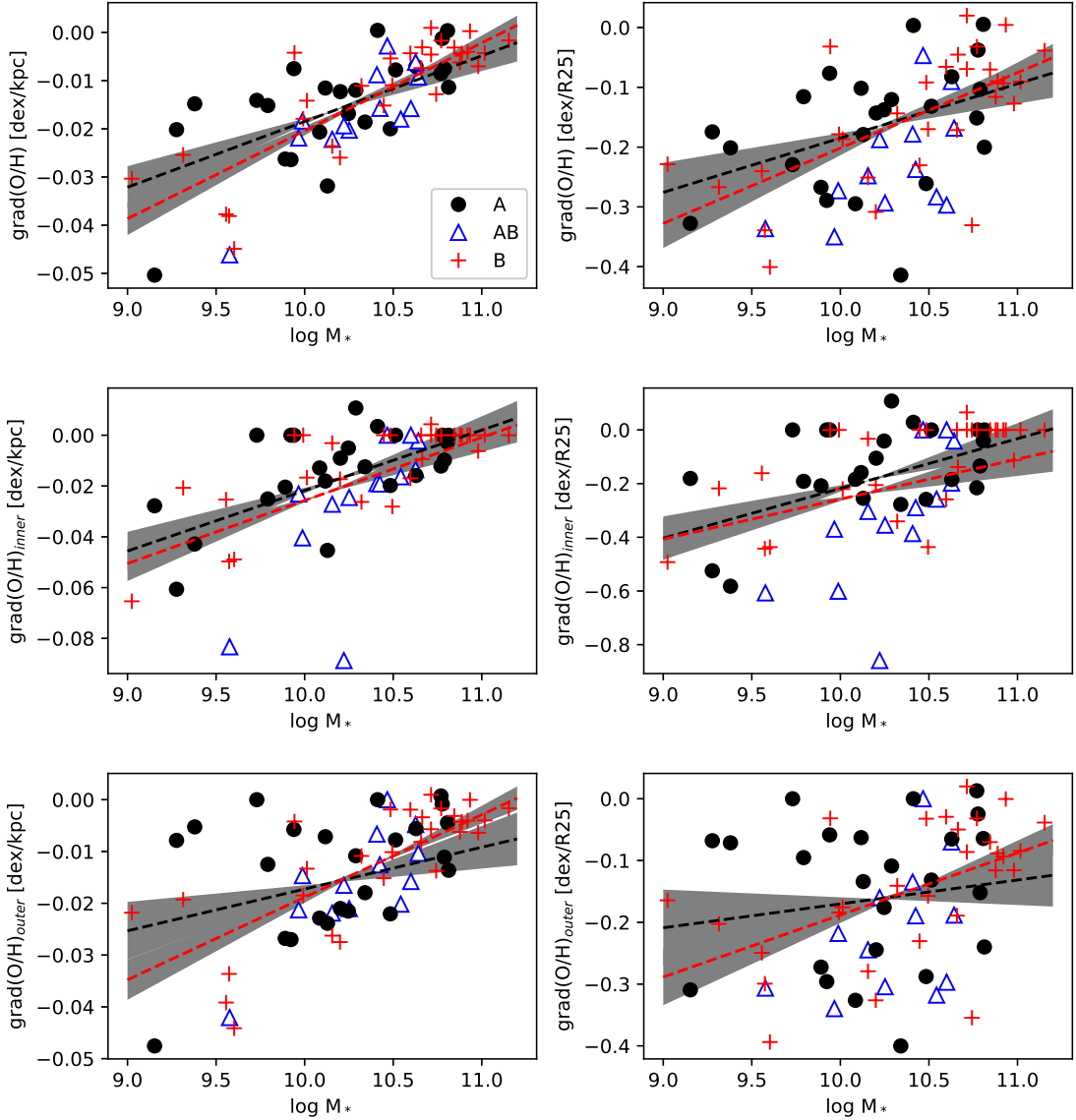}
\end{center}
\caption{Stellar mass -- oxygen abundance gradient diagrams for the abundance gradient scaled to 
kpc (left columns) and $R_{25}$ (right columns). Top panels: Oxygen abundance gradient for galactocentric 
distances $r < R_{25}$. Middle panels: Oxygen abundance gradient for $r < 0.25R_{25}$. Bottom panels:
Oxygen abundance gradient for $0.25 < r < R_{25}$. 
Unbarred (A), barred (B), and intermediate (AB) galaxies are shown with different colors and symbols.
Black dashed lines are the best fit for the subsample 
of unbarred galaxies, and red lines show this for the barred galaxies. Shaded areas represent the 1$\sigma$ 
uncertainties of the fit.
}
\label{figure:grad-M-bar}
\end{figure*}

\section{Oxygen abundance gradient}

We fit the radial oxygen abundance distribution in the disk using the 
following relation: 
\begin{equation}
12+\log({\rm O/H})  = 12+\log({\rm O/H})_{0} + grad \times R ,
\label{equation:grad}
\end{equation} 
where 12 + log(O/H)$_{0}$ is the oxygen abundance at $R_{0} = 0$, that is, the extrapolated 
value of the oxygen abundance in the galactic center, $grad$ is the slope of the oxygen abundance gradient.
It is known that the oxygen abundances in the inner and outer parts of some galaxies are systematically different 
from those expected from the overall abundance gradient 
\citep[e.g.,][]{Belley1992,Bresolin2009,Goddard2011,RosalesOrtega2011,Werk2011,Bresolin2012,Patterson2012,Sanchez2012,Sanchez2014,SanchezMenguiano2016,SanchezMenguiano2018,Zinchenko2016}.
However, within an optical radius, these differences are relatively small, smaller than 0.05~dex, and a
break in the slope of the radial oxygen abundance distribution can be found at any radius \citep{Pilyugin2017}.
Thus, the oxygen abundance gradient was estimated in three ranges of galactocentric distances: 
within $R_{25}$, in the inner $R < 0.25 R_{25}$, and outer $0.25 R_{25} < R < R_{25}$ parts 
of the galaxy. It is presented in Table~\ref{table:sample}
\footnote{A detailed list of oxygen abundance gradients and errors for each galaxy is available 
at https://sites.google.com/view/igorzinchenko/main/califa}.

Figure~\ref{figure:NGC0257-grad} shows the oxygen abundance gradient for NGC~257. Circles represent the oxygen abundance 
in the individual spaxels. 
The black solid line is the best fit to spaxels within the optical radius $R_{25}$ of the galaxy.
Dashed red lines correspond to the best fit of the oxygen abundance gradient in the 
inner and outer parts of the galaxy. The radial gradient flattening is clearly seen in the inner part of the galaxy.

Figure~\ref{figure:grad-M-bar} shows the oxygen abundance gradient as a function of stellar mass. 
Black circles, red plus signs, and blue triangles represent unbarred (A), barred (B), and intermediate (AB) galaxies, 
respectively. Black dashed lines are the best fit for the subsample of unbarred galaxies, and red 
lines show this for the barred galaxies. Shaded areas represent the 1$\sigma$ uncertainties of the fit. 
The left panels show the abundance gradient scaled to kiloparsecs, while the gradient in the right panels 
is scaled to $R_{25}$. In all cases there is no significant difference between barred and 
unbarred galaxies. 

Galaxies with higher mass exhibit flatter oxygen abundance gradients 
when the gradient is measured per kpc (left panels of the Figure~\ref{figure:grad-M-bar}),
in agreement with the result of the numerical simulation of \citet{Tissera2016}. 
This effect is significantly less prominent for the oxygen abundance gradients scaled to $R_{25}$ 
(right panels of the Figure~\ref{figure:grad-M-bar}) and almost disappears for the gradient 
calculated in the outer region (bottom right panel). Our result confirms the 
conclusion of \citet{Ho2015}, who attributed to the size effect
the different dependence of the oxygen abundance
gradient on the stellar mass for the oxygen abundance gradient normalized to kpc or $R_{25}$. 
They noted that galaxies with steeper oxygen abundance gradients 
usually have a smaller optical radius. Thus, our result suggests that at the present epoch, 
the variation in oxygen abundance between the central and the outer parts of galaxies is similar for galaxies 
in the whole mass range, while the radial variation in absolute scale (e.g., per kpc) is 
larger for less massive galaxies.


\begin{figure*}
\begin{center}
\includegraphics[width=0.85\linewidth]{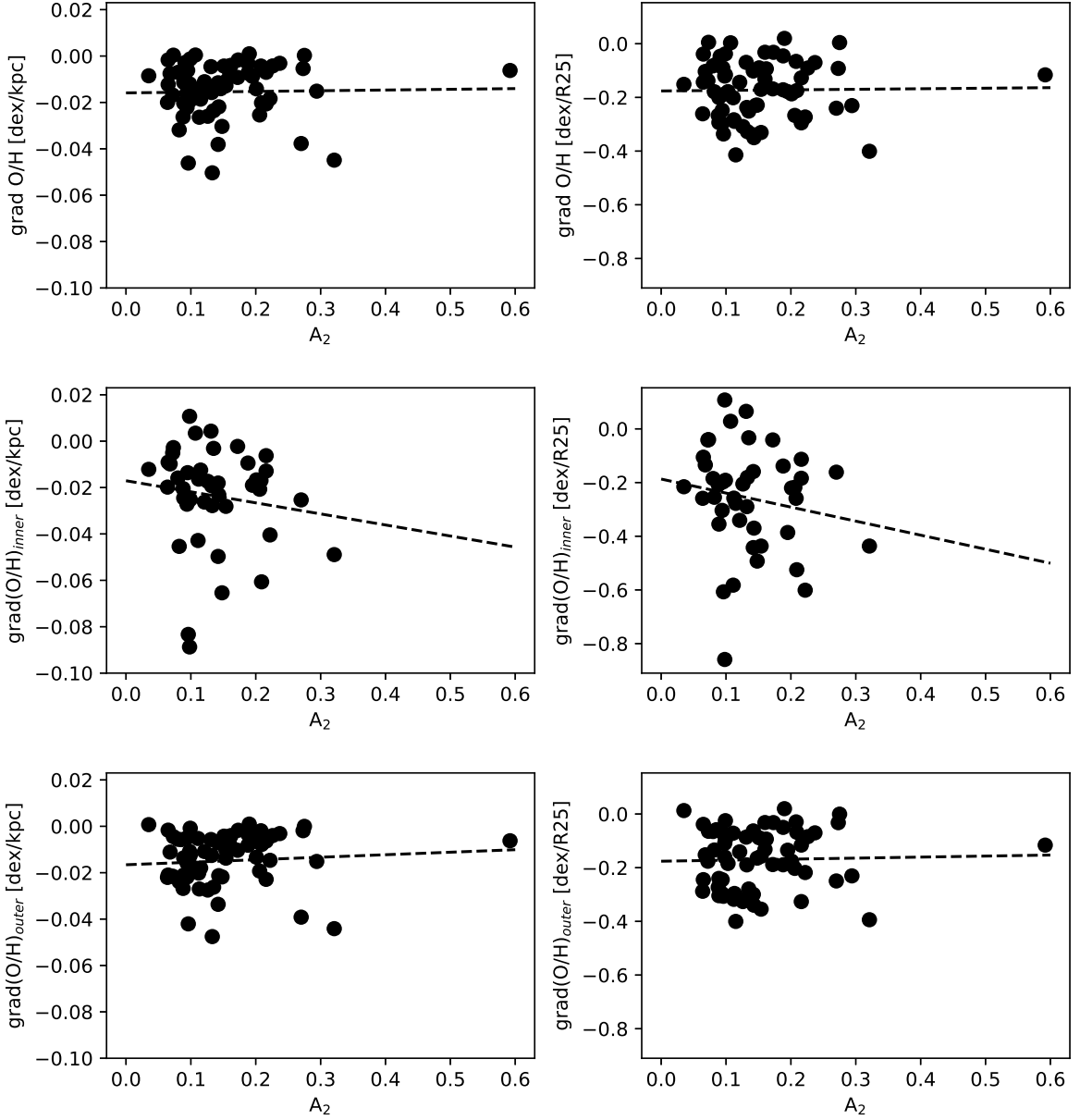}
\end{center}
\caption{Diagrams showing the A$_2$ Fourier coefficient -- oxygen abundance gradient for the abundance gradient scaled to 
kpc (left columns) and $R_{25}$ (right columns). Top panels: Oxygen abundance gradient for galactocentric 
distances $r < R_{25}$. Middle panels: Oxygen abundance gradient for $r < 0.25R_{25}$. Bottom panels:
Oxygen abundance gradient for $0.25 < r < R_{25}$.
}
\label{figure:grad-A2-bar}
\end{figure*}

Different approaches to quantify the strength of the bar based on
the axis ratio and bar length \citep{Martin1995,Martinet1997,Chapelon1999,Aguerri1999,Abraham2000,Marinova2007},
bar torque \citep{Combes1981,Buta2003,Laurikainen2002,Buta2003,DiazGarcia2016},
and Fourier analysis \citep{Aguerri1998,Aguerri2000,DiazGarcia2016,GarciaGomez2017}, for instance, 
have been suggested and used in the literature.
Following \citet{DiazGarcia2016} and \citet{Saha2013}, 
to quantify the strength of the non-axisymmetric structures such as the bar and/or spiral patterns
in the disk of a galaxy, we calculated the discrete Fourier transform of the surface brightness 
in the SDSS $r$ band in the disk 
of each galaxy. After correcting for inclination, we divided the galactic disk into concentric rings 
and defined the Fourier amplitude coefficient in each ring. The $m$-th complex Fourier coefficient is 
given by 
\begin{equation} 
\tilde{\rm A}_m({\rm R_k}) = \sum_{j=0}^{{\rm N}-1}~ {\rm I}_j ~e^{-2\pi i ~m ~j/{\rm N}} 
,\end{equation} 
where I$_j$ is the surface brightness within the $j$-th pixel in a given ring R$_k$. 
$A_m = \tilde{A}_m/\tilde{A}_0$ are the normalized Fourier amplitudes. $\tilde{A}_0$
refers to the axisymmetric surface brightness component (mean surface brightness at the 
given radius), and $\tilde{A}_m$ is an amplitude
of the $m$-th mode of the azimuthal variation of surface brightness. 
For the further analysis, we adopted a maximum value of the $A_2$ Fourier amplitude 
for a galactocentric radius smaller than $0.5R_{25}$, which corresponds to the maximum 
length of a bar \citep{Erwin2005}. These derived maxima of the $A_2$ Fourier amplitudes for each galaxy are presented in Table~\ref{table:sample}.
We did not consider Fourier amplitudes in the center of galaxies ($R < 2$ arcsec) either because they
can be affected by the uncertainties in the position of the galactic center.

Figure~\ref{figure:grad-A2-bar} shows the oxygen abundance gradient as a function of the 
maximum value of the $A_2$ Fourier amplitude. The dashed line is the linear fit of the 
data. We applied a two-sided p-value statistical test whose null hypothesis is that the slope of 
the $grad$ versus $A_2$ relation is zero.
The p-values range from 0.30 to 0.86. We thus confirmed the null hypothesis of no correlation 
between oxygen abundance gradient and the $A_2$ Fourier amplitude of the surface brightness 
in the SDSS $r$ band.

\section{Generalized mass -- metallicity relation}

The oxygen abundance in a galaxy can depend on the galaxy mass \citep{Tremonti2004,Kewley2008,Thuan2010}, 
the surface brightness \citep{Vila-Costas1992,Ryder1995,Pilyugin2014b},
the disk scale length \citep{Pilyugin2014b,Zinchenko2015},
the morphological type \citep{Pilyugin2014b,Zinchenko2015}, 
and the galactocentric radius \citep[among many others]{Peimbert1979,Shields1978,Sanchez2014,Zinchenko2016,Belfiore2017}. 
Moreover, the dependence on the given parameter can be seen 
only at some range of galactocentric distances \citep{Pilyugin2014b}. We therefore considered the generalized 
mass -- metallicity relation as a multiparameter function to distinguish the impact of size, color,
disk surface brightness, and strength of bar/spirals on the oxygen abundance in the central, intermediate, and 
outer parts of the galaxy.

For a given galactocentric radius we applied the following parametric relation: 
\begin{eqnarray}
       \begin{array}{lll}
12+\log {\rm (O/H)}    & = &  c_0 + c_1 \, \log M_* + c_2 \, (\log M_*)^{2}  \\  
                       & + & c_{R25} \, R_{25} + c_{g-r} \, (g-r) \\
                       & + & c_{\Sigma} \, \log \Sigma_0 + c_{A2} \, A_2\\ 
     \end{array}
\label{equation:oh-mass}
,\end{eqnarray}
where $M_*$ is the stellar mass of the galaxy expressed in the mass of the Sun,
$R_{25}$ is the isophotal radius of a galaxy in kpc, $g-r$ is the difference
between the total magnitude of the galaxy in the SDSS $g$ and $r$ bands, $\Sigma_0$ is the 
central surface brightness of the disk in the SDSS $r$ band  
in units of $L_{\odot}$/pc$^2$, and $A_2$ is the maximum
value of the $A_2$ Fourier coefficient. This relation was constructed for the three 
galactocentric radii of $0, 0.5 R_{25}$, and $R_{25}$. The oxygen abundance at a given 
radius was calculated using the fit of the abundance gradient. Abundances
at the center of each galaxy were obtained from the inner region gradient fit ($R < R_{25}$), 
while oxygen abundances at $R = 0.5 R_{25}$ and  $R = R_{25}$ were estimated using 
the outer gradient fit ($0.25R_{25} < R < R_{25}$). The values of each coefficient for 
the three considered radii are listed in Table~\ref{table:MZparams}. 

The disk surface brightness was obtained in the following way. We corrected the surface brightness
of each pixel for Galactic foreground extinction using the given in the NASA 
Extragalactic Database (NED) \footnote{https://ned.ipac.caltech.edu/}. We applied the $A_r$ 
values from the recalibration by \citet{Schlafly2011} of the \citet{Schlegel1998} 
infrared-based dust map. Then, we fit the radial surface brightness profile by 
an exponential profile 
\begin{equation}
\Sigma(R) = \Sigma_0 \, \exp(-R/h)
\label{equation:exp-profile}
,\end{equation}
where $\Sigma_0$ is the central disk surface brightness and $h$ is the radial scale length.
As the radial surface brightness profiles of many galaxies are affected by the bulge in the 
central part, we fit the surface brightness for the galactocentric distances $0.3R_{25} < R < R_{25}$.
The surface brightness was converted into solar units for the further analysis. 
The magnitude of the Sun in the SDSS $r$ band $M_{\odot,r} = 4.64$ was taken from \citet{Blanton2007}.

\begin{figure*}
\begin{center}
\includegraphics[width=1.0\linewidth]{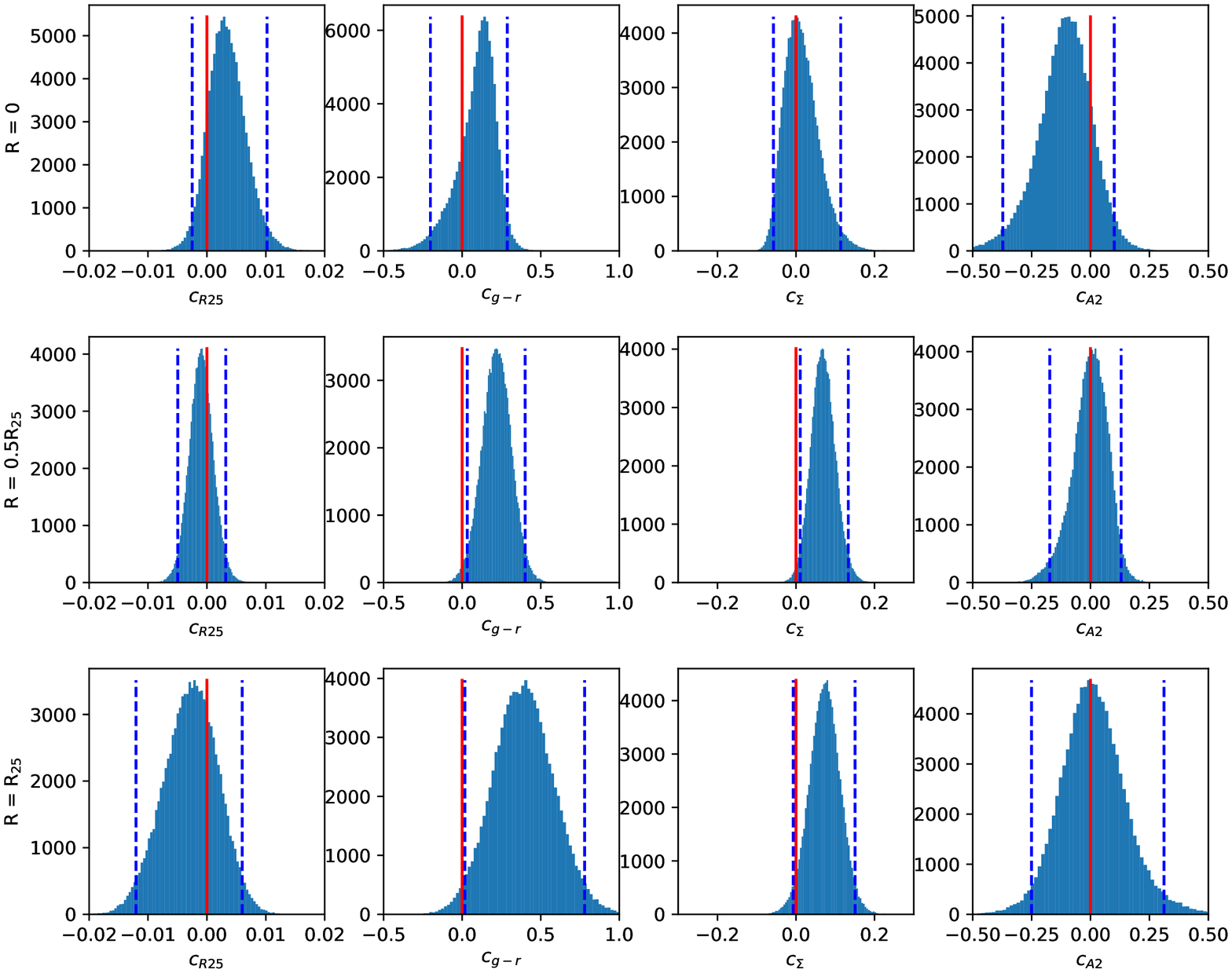}
\end{center}
\caption{Probability distribution function for the $c_{R25}$, $c_{g-r}$, $c_{\Sigma}$, and $c_{A2}$ coefficients 
of the generalized mass -- metallicity relation in the center of galaxies (top panels), at $R = 0.5R_{25}$ (middle
panels), and $R = R_{25}$ (bottom panels). The solid vertical line represents the zero value 
for each coefficient, and the dashed lines show the two-sided 95\% confidence interval. 
}
\label{figure:hist-coef-OH_break_A2}
\end{figure*}

We estimated the probability distribution function (PDF) and the 95\% confidence intervals for 
the $c_{R25}$, $c_{g-r}$, $c_{\Sigma}$, and $c_{A2}$ coefficients of the generalized mass -- metallicity 
relation using a bootstrap method with 100,000 iterations (Figure~\ref{figure:hist-coef-OH_break_A2}).  
We considered the generalized mass -- metallicity relation for the local oxygen abundance 
in the center, at the half of the optical radius, and at the optical radius of a galaxy.
For the central part of a galaxy, the zero value falls within the 95\% confidence interval of the PDF for all 
parameters. This shows that the oxygen abundance in the central part of a galaxy depends  
neither on the optical radius $R_{25}$ nor on the color $g-r$, $\Sigma_0$, and $A_2$ of the galaxy.
However, outside the central part, the oxygen abundance increases with $g-r$ and $\Sigma_0$
at a confidence level of almost 95$\%$. 
The right side columns of Figure~\ref{figure:hist-coef-OH_break_A2} show that there is no dependence 
of the generalized mass--metallicity relation on the $A_2$ parameter because the zero value falls 
within the 95\% confidence interval of the PDF at all radii.

\begin{figure*}
\begin{center}
\includegraphics[width=1.0\linewidth]{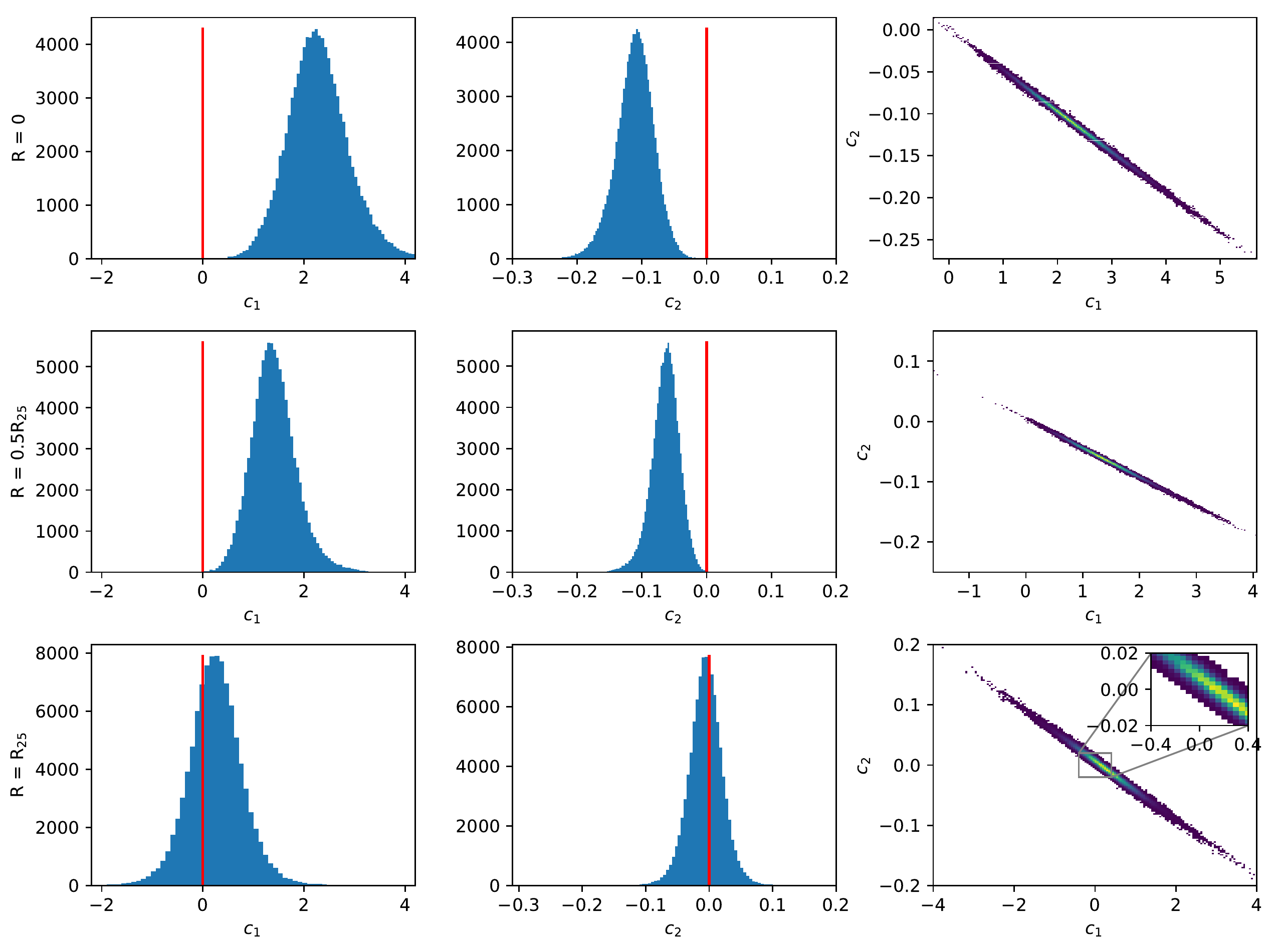}
\end{center}
\caption{Probability distribution function for the $c_1$ (left column) and $c_2$ (middle column) coefficients 
of the generalized mass -- metallicity relation in the center of galaxies (top panels), at $R = 0.5R_{25}$ (middle
panels), and $R = R_{25}$ (bottom panels). The solid vertical line represents the zero value 
for each coefficient. The right column represents the correlation between the $c_1$ and $c_2$ coefficients.
}
\label{figure:hist-coef-mass}
\end{figure*}

Figure~\ref{figure:hist-coef-mass} shows the probability distribution function for the $c_1$ (left column) 
and $c_2$ (middle column) coefficients of the generalized mass -- metallicity relation in the center of galaxies (top panels), 
at $R = 0.5R_{25}$ (middle panels), and $R = R_{25}$ (bottom panels). The solid vertical line represents the zero value 
for each coefficient. The coefficient $c_2$ represents the nonlinearity of the generalized mass -- metallicity relation,
which is significant in the inner region of galaxies. 
The right column panels, which are 2D PDFs 
in $c_1$ and $c_2$ coordinate axes, represent the strong correlation between the $c_1$ and $c_2$ 
coefficients at all galactocentric radii. Because of this and the short mass range, the width of the PDF of $c_1$ and $c_2$ 
coefficients is quite large.
For the outer part of the galaxy, the PDF centers for the $c_1$ and $c_2$ coefficients are close to zero. 
This might be thought to indicate that in the outer parts of the galaxies the oxygen abundance does not depend on the galaxy mass.
To determine whether this is true, we examined the right bottom panel, which represents the correlation between 
$c_1$ and $c_2$ for the outer region of a disk. This panel contains the zoomed central part of the distribution. 
It is very unlikely that the $c_1$ and $c_2$ coefficients are both equal to zero simultaneously.
This means that the oxygen abundance depends on the galaxy mass at the outer part of a galaxy as well.

We here limit our sample to face-on galaxies of large angular size in order to resolve 
the bar and spiral structure. 
In a future work we will investigate the nonlinearity of the generalized mass--metallicity relation
in detail using a larger galaxy sample.

\begin{table*}
\caption[]{\label{table:sample}
Adopted and derived properties of our target galaxies
}
\begin{center}
\begin{tabular}{lrrrrrrrcccccccc} \hline \hline
Name      & D$^a$  &Type$^b$& $PA$ &  $i$   & $\log M$$^b$ & $R_{25}$& $g-r$  & $\log \Sigma_0$& $A_2$ & grad(O/H) & grad(O/H)$_{inn}$ & grad(O/H)$_{out}$   \\ 
          & Mpc    &       & deg  &  deg   & $M_\odot$& kpc     &  mag   & $L_\odot/pc^2$&        & dex/$R_{25}$  & dex/$R_{25}$  & dex/$R_{25}$          \\ \hline
NGC~1     &  61.6  &  SAbc & 102  &  41.4  &  10.629  &  11.65  &  0.517  &  2.339  &    0.080 & -0.082 & -0.185 & -0.065  \\
NGC~23    &  61.7  &  SBb  & 155  &  59.3  &  10.980  &  18.13  &  0.514  &  2.420  &    0.216 & -0.127 & -0.113 & -0.116 \\
NGC~171   &  52.8  &  SBb  &  98  &  25.8  &  10.447  &  15.21  &  0.454  &  2.586  &    0.294 & -0.230 &    --- & -0.230  \\
NGC~180   &  70.6  &  SBb  & 159  &  50.2  &  10.658  &  23.21  &  0.395  &  2.157  &    0.188 & -0.172 &    --- & -0.189   \\
NGC~214   &  61.0  &  SABbc&  58  &  46.4  &  10.467  &  16.68  &  0.383  &  2.586  &    0.091 & -0.047 &    --- &  ---   \\
NGC~234   &  59.7  &  SABc &  71  &  29.5  &  10.627  &  14.59  &  0.389  &  2.702  &    0.095 & -0.090 & -0.198 & -0.070  \\
NGC~237   &  55.9  &  SBc  & 178  &  53.1  &  10.200  &  11.87  &  0.390  &  2.466  &    0.126 & -0.308 & -0.205 & -0.326  \\
NGC~257   &  70.4  &  SAc  &  94  &  52.4  &  10.813  &  17.61  &  0.462  &  2.494  &    0.089 & -0.200 &  0.000 & -0.240  \\
NGC~309   &  75.8  &  SBcd & 112  &  25.8  &  10.743  &  25.80  &  0.297  &  2.450  &    0.154 & -0.331 &    --- & -0.355  \\
NGC~477   &  89.5  &  SABbc& 138  &  52.4  &  10.408  &  20.31  &  0.379  &  2.205  &    0.195 & -0.178 & -0.386 & -0.135 \\
NGC~776   &  65.5  &  SBb  & 136  &  38.7  &  10.597  &  15.24  &  0.383  &  2.368  &    0.208 & -0.066 & -0.259 & -0.029 \\
NGC~941   &  21.3  &  SAcd & 175  &  28.4  &   9.153  &   6.51  &  0.175  &  2.155  &    0.133 & -0.328 & -0.181 & -0.309 \\
NGC~976   &  57.4  &  SAbc & 164  &  39.6  &  10.789  &  13.69  &  0.445  &  2.574  &    0.068 & -0.103 & -0.134 & -0.152 \\
NGC~991   &  20.2  &  SABcd& 126  &  18.2  &   9.577  &   7.29  &  0.320  &  2.254  &    0.096 & -0.336 & -0.607 & -0.307 \\
NGC~1070  &  54.0  &  SAb  &   2$^c$  &  33.9$^c$  &  10.770  &  17.75  &  0.486  &  2.384  &    0.035 & -0.151 & -0.216 &  0.013 \\
NGC~1094  &  85.8  &  SABb &  95  &  56.6  &  10.642  &  18.22  &  0.419  &  2.120  &    0.172 & -0.168 & -0.041 & -0.188\\
NGC~1659  &  61.7  &  SABbc&  49  &  51.0  &  10.425  &  15.08  &  0.387  &  2.560  &    0.132 & -0.237 & -0.290 & -0.190\\
NGC~1667  &  61.2  &  SBbc & 172  &  45.6  &  10.714  &  15.13  &  0.372  &  2.960  &    0.131 & -0.069 &  0.065 & -0.086\\
NGC~2347  &  63.0  &  SABbc&   3  &  54.6  &  10.543  &  15.76  &  0.412  &  2.409  &    0.112 & -0.283 & -0.258 & -0.318\\
NGC~2487  &  68.9  &  SBb  &  44  &  44.8  &  10.714  &  20.84  &  0.449  &  2.255  &    0.190 &  0.020 &    --- &  0.020\\
NGC~2530  &  73.0  &  SABd & 152  &  34.9  &   9.988  &  14.86  &  0.232  &  2.160  &    0.222 & -0.273 & -0.601 & -0.217\\
NGC~2540  &  89.0  &  SBbc & 126  &  47.9  &  10.495  &  15.53  &  0.433  &  2.384  &    0.154 & -0.170 & -0.437 & -0.157\\
NGC~2604  &  32.3  &  SBd  & 174  &  19.9  &   9.602  &   8.93  &  0.202  &  2.267  &    0.321 & -0.401 & -0.437 & -0.394\\
NGC~2730  &  56.7  &  SBcd &  72  &  45.6  &  10.013  &  13.19  &  0.298  &  2.248  &    0.201 & -0.187 & -0.220 & -0.176\\
NGC~2805  &  28.7  &  SAc  & 175  &  30.7  &   9.730  &  16.28  &  0.134  &  1.569  &    0.146 & -0.229 &    --- &    ---\\
NGC~2906  &  33.5  &  SAbc &  82  &  56.6  &  10.412  &   8.19  &  0.352  &  2.741  &    0.107 &  0.004 &  0.028 &  0.000\\
NGC~2916  &  56.0  &  SAbc &  16  &  49.5  &  10.514  &  16.94  &  0.406  &  2.628  &    0.160 & -0.132 &    --- & -0.132\\
NGC~3057  &  25.9  &  SBdm &   2  &  56.6  &   9.024  &   7.53  &  0.134  &  1.741  &    0.148 & -0.229 & -0.493 & -0.164\\
NGC~3381  &  22.8  &  SBd  &  40  &  33.9  &   9.557  &   6.37  &  0.264  &  2.334  &    0.270 & -0.240 & -0.161 & -0.249\\
NGC~3614  &  38.4  &  SABbc& 100  &  50.2  &   9.966  &  15.97  &  0.307  &  2.044  &    0.143 & -0.350 & -0.370 & -0.340\\
NGC~3687  &  41.1  &  SBb  & 150  &  23.1  &  10.156  &  10.64  &  0.437  &  2.485  &    0.135 & -0.251 & -0.033 & -0.279\\
NGC~3811  &  49.0  &  SBbc &  14  &  48.7  &  10.321  &  12.97  &  0.426  &  2.530  &    0.121 & -0.144 & -0.341 & -0.141\\
NGC~4185  &  61.0  &  SABbc& 167  &  47.9  &  10.600  &  18.81  &  0.496  &  2.279  &    0.094 & -0.297 &    --- & -0.297\\
NGC~4961  &  42.5  &  SBcd & 111  &  50.9  &   9.574  &   8.90  &  0.236  &  2.173  &    0.142 & -0.339 & -0.442 & -0.299\\
NGC~5000  &  84.9  &  SBbc &  69  &  48.7  &  10.485  &  17.04  &  0.420  &  2.308  &    0.273 & -0.092 &    --- & -0.032\\
NGC~5016  &  43.5  &  SAbc &  58  &  46.4  &  10.202  &  11.64  &  0.433  &  2.503  &    0.065 & -0.143 & -0.105 & -0.245\\
NGC~5205  &  30.9  &  SBbc & 157  &  53.8  &   9.943  &   7.55  &  0.451  &  2.385  &    0.160 & -0.032 &    --- & -0.032\\
NGC~5320  &  34.3  &  SABbc&  14  &  58.0  &  10.156  &  11.17  &  0.340  &  2.316  &    0.094 & -0.248 & -0.304 & -0.245\\
NGC~5406  &  79.0  &  SBb  & 106  &  29.5  &  11.017  &  21.14  &  0.676  &  2.612  &    0.226 & -0.091 &    --- & -0.084\\
NGC~5480  &  32.9  &  SAcd &  11  &  42.3  &  10.117  &   8.80  &  0.462  &  2.759  &    0.142 & -0.102 & -0.159 & -0.063\\
NGC~5520  &  33.2  &  SAbc &  65  &  58.7  &   9.792  &   7.63  &  0.370  &  2.346  &    0.099 & -0.116 & -0.191 & -0.095\\
NGC~5633  &  39.6  &  SAbc &  14  &  48.7  &  10.247  &   8.18  &  0.487  &  2.969  &    0.072 & -0.138 & -0.041 & -0.176\\
NGC~5720  & 113.1  &  SBbc & 130  &  48.7  &  10.845  &  22.37  &  0.458  &  2.256  &    0.237 & -0.070 &    --- & -0.070\\
NGC~5732  &  59.3  &  SAbc &  40  &  54.5  &   9.891  &  10.18  &  0.402  &  2.190  &    0.088 & -0.267 & -0.208 & -0.273\\
NGC~5888  & 126.7  &  SBb  & 158  &  56.0  &  11.153  &  23.22  &  0.626  &  2.408  &    0.065 & -0.039 &    --- & -0.039\\
NGC~5957  &  32.0  &  SBb  &  91  &  38.7  &   9.992  &   9.96  &  0.321  &  2.544  &    0.103 & -0.178 &    --- & -0.184\\
NGC~6004  &  60.8  &  SBbc &  93  &  20.0  &  10.664  &  14.68  &  0.482  &  2.522  &    0.188 & -0.045 & -0.138 & -0.050\\
NGC~6063  &  46.7  &  SAbc & 154  &  56.0  &   9.939  &  10.19  &  0.360  &  2.312  &    0.085 & -0.077 &    --- & -0.059\\
NGC~6154  &  88.7  &  SBab & 137  &  48.7  &  10.771  &  18.84  &  0.552  &  2.179  &    0.173 & -0.032 &    --- & -0.032\\
NGC~6155  &  26.1  &  SAc  & 147  &  45.6  &  10.129  &   5.62  &  0.471  &  2.723  &    0.082 & -0.179 & -0.255 & -0.134\\
NGC~6301  & 120.3  &  SAbc & 109  &  53.1  &  10.777  &  30.09  &  0.380  &  2.222  &    0.099 & -0.038 &    --- & -0.025\\
NGC~6497  &  84.8  &  SBab & 116  &  57.3  &  10.934  &  16.53  &  0.613  &  2.324  &    0.275 &  0.004 &    --- & -0.000\\
NGC~6941  &  88.6  &  SBb  & 125  &  46.4  &  10.920  &  23.20  &  0.509  &  2.213  &    0.162 & -0.093 &    --- & -0.093\\
NGC~7321  &  97.9  &  SBbc &  21  &  50.2  &  10.888  &  20.79  &  0.457  &  2.443  &    0.151 & -0.090 &    --- & -0.088\\
NGC~7364  &  67.1  &  SAab &  64  &  51.7  &  10.807  &  14.44  &  0.521  &  2.292  &    0.073 &  0.005 & -0.040 & -0.064\\
NGC~7489  &  85.1  &  SAbc & 164  &  56.6  &  10.341  &  22.28  &  0.097  &  1.968  &    0.115 & -0.414 & -0.278 & -0.400\\
NGC~7653  &  58.3  &  SAb  & 168  &  31.8  &  10.484  &  13.06  &  0.399  &  2.375  &    0.064 & -0.261 & -0.259 & -0.288\\
NGC~7716  &  35.6  &  SAb  &  31  &  39.6  &  10.289  &  10.04  &  0.494  &  2.457  &    0.098 & -0.120 &  0.108 & -0.109\\
NGC~7738  &  91.4  &  SBb  &   5  &   0.0  &  10.877  &  18.61  &  0.618  &  1.804  &    0.592 & -0.116 &    --- & -0.116\\
NGC~7819  &  67.2  &  SAc  &  88  &  53.1  &  10.085  &  14.27  &  0.312  &  1.884  &    0.216 & -0.295 & -0.184 & -0.326\\
IC~776    &  40.2  &  SAdm &  92  &  53.8  &   9.277  &   8.65  &  0.202  &  1.590  &    0.209 & -0.174 & -0.525 & -0.068\\
IC~1256   &  72.1  &  SABb &  93  &  51.7  &  10.252  &  14.47  &  0.381  &  2.288  &    0.089 & -0.294 & -0.355 & -0.304\\
IC~5309   &  57.4  &  SABc &  26  &  59.3  &  10.222  &   9.68  &  0.487  &  2.043  &    0.098 & -0.188 & -0.859 & -0.160\\
UGC~8733  &  39.7  &  SBdm &   6  &  58.0  &   9.314  &  10.51  &  0.196  &  1.701  &    0.206 & -0.267 & -0.218 & -0.203\\
UGC~12224 &  43.3  &  SAc  &  36  &  30.7  &   9.923  &  10.96  &  0.290  &  2.077  &    0.113 & -0.289 &    --- & -0.296\\
UGC~12816 &  71.9  &  SAc  & 142  &  54.5  &   9.379  &  13.59  &  0.167  &  1.822  &    0.111 & -0.201 & -0.582 & -0.071\\
\hline 
\end{tabular}\\
\end{center}
\begin{flushleft}
$^{a}$ https://ned.ipac.caltech.edu/ \\
$^{b}$ \citet{CALIFA2014} \\
$^{c}$ derived from Sersic profile
\end{flushleft}

\end{table*}

\begin{table*}
\caption[]{\label{table:MZparams}
Coefficients of parametric mass--metallicity relation.
}
\begin{center}
\begin{tabular}{rccccccc} \hline \hline
      R      & $c_0$  & $c_1$   &  $c_2$ & $c_{R25}$ & $c_{g-r}$ & $c_{\Sigma}$ & $c_{A2}$ \\ \hline
$0$          & -2.7417 & 2.192 & -0.106 &  0.004 & 0.133 & 0.017 & -0.093 \\
$0.5 R_{25}$ &  0.7301 & 1.377 & -0.062 & -0.001 & 0.222 & 0.071 &  0.005 \\
$R_{25}$     &  6.2023 & 0.245 & -0.005 & -0.002 & 0.397 & 0.074 & -0.005 \\ \hline
\end{tabular}\\
\end{center}
\end{table*}

\section{Discussion and summary} 

We investigated the possible dependence of the oxygen abundance gradient in 
disk galaxies on the presence of a bar/spiral and other parameters of the galaxy, such as 
mass, size, color index, and surface brightness. We also studied the generalized mass--metallicity relation 
for the oxygen abundance in the center, intermediate, and outer parts of the galaxy
as a multivariable relation including size, color index, surface brightness, and the bar/spiral 
pattern strength. Our sample contains 66 galaxies from the CALIFA DR3 survey.
We adopted the Fourier amplitude A$_2$ of the surface brightness as the quantitative characteristic 
of the strength of non-axisymmetric structures in a galactic disk, in addition to 
a commonly used morphologic division into A, AB, and B types based on the Hubble classification.

We considered the generalized mass--metallicity relation in order to take into account a possible 
dependence of the local oxygen abundance and the strength of the non-axisymmetric structures in a disk.
Our relation also takes into account the influence of
the isophotal radius $R_{25}$, the color index $g-r$ in the SDSS bands, the central surface brightness of the disk 
$\Sigma$ in the SDSS $r$ band, and the bar/spiral strength, quantified as the maximum
amplitude of the $A_2$ Fourier coefficient. This approach allowed us to distinguish changes in local 
oxygen abundance that are caused by each parameter even when the oxygen abundance depends on many 
parameters simultaneously. This relation was constructed for the three 
galactocentric radii of $0, 0.5 R_{25}$, and $R_{25}$.

We find larger negative oxygen abundance gradients scaled to kpc for the galaxies with lower masses.
However, this effect is significantly less prominent for the oxygen abundance gradients scaled to $R_{25}$ 
, as has been shown by previous observational \citep{Ho2015} and theoretical \citep{Prantzos2000} studies.
The dependence of oxygen abundance gradient on the stellar mass almost disappears when 
the gradient is calculated for the outer part of the galaxy and scaled to $R_{25}$.

We find neither a significant difference between the oxygen abundance gradient in unbarred 
and barred galaxies (Hubble types A an B) nor a dependence of the mass--local metallicity relation on the 
$A_2$ parameter at any galactocentric radius. A standard deviation of the difference between 
oxygen abundances, observed and calculated with our generalized mass--metallicity relation, ranges 
from 0.04~dex at the center to 0.07~dex at the optical radius. Thus, our data suggest that 
there is no significant impact of the non-axisymmetric structures such as a bar and/or spiral patterns
on the oxygen abundance and radial metallicity gradient of the spiral galaxies. The 
effect of the oxygen abundance distribution is smaller than $\sim 0.05$~dex at any distance from the center of a galaxy. 
This conclusion agrees with 
the results on the stellar \citep{Sanchez-Blazquez2014,Cheung2015} and gas-phase \citep{SanchezMenguiano2016} 
metallicity gradients as well as 
with a few recent studies of the azimuthal variation in oxygen abundance 
(in the spiral arms and interarm medium) for the CALIFA galaxies. \citet{Sakhibov2018} showed that the enhancement of the 
oxygen abundance in the spiral arms as compared to the interarm regions is lower that 0.05~dex. 
Moreover, no significant changes in radial oxygen abundance gradient between the spiral arms and 
the interarm medium have been found. In addition, \citet{Sanchez-Menguiano2017} noted some 
differences in the azimuthal distribution of the oxygen abundance between the arm and interarm star-forming regions in barred 
and flocculent spirals. They suggested that this influence is not strong enough to affect the overall abundance distribution,
which is confirmed by our results.

The absence of a correlation between the gas-phase oxygen abundance and the non-axisymmetric structures in galaxy disks suggests that at least for the oxygen, 
the processes of chemical enrichment of the interstellar medium are more efficient 
in producing the radial abundance gradient than the  gas mixing. 

We find that the local oxygen abundance in the outer part of the disk increased with the 
central surface brightness of the disk, while there is no such correlation for the central part 
of the galaxies. This result is consistent with the conclusion of \citet{Wu2015}. They considered 
the averaged oxygen abundance in nearby galaxies with respect to its surface brightness and found 
that galaxies with lower surface brightness tend to have lower metallicity.
The correlation between local oxygen abundance and surface brightness,
in addition to the flattening of the radial metallicity gradient in 
the massive galaxies, can be considered as confirmation of the inside-out galaxy formation 
scenario \citep{Kepner1999,Pilkington2012}. If the chemical evolution in the 
central part of the galaxy is almost completed, the dependence of the metallicity on the galaxy 
parameters should indeed be suppressed.

Our findings are briefly summarized below.

\begin{enumerate}
 \item We showed that galaxies with higher stellar mass exhibit flatter oxygen abundance gradients. 
This correlation becomes more prominent for gradients expressed in units of dex/kpc 
and obtained for the central part of the galaxies ($R < 0.25R_{25}$) and for the galactocentric 
distances in the range from $0$ to $R_{25}$. 
  
 \item There is no significant difference between the oxygen abundance gradient in unbarred 
 and barred galaxies (Hubble types A and B). In addition, we did not find a dependence of the 
 generalized mass--metallicity relation on the 
$A_2$ parameter. Our data suggest that there is no significant impact of the 
non-axisymmetric structures such as a bar and/or spiral patterns on the local oxygen abundance
in the disk of a galaxy or this impact is weaker than $\sim 0.05$~dex. 
However, this correlation is significantly reduced when the abundance 
gradients are scaled to $R_{25}$, and it almost disappears when the gradients are calculated for the outer part 
of the galaxy.
 
 \item The local oxygen abundance does not depend on the optical radius $R_{25}$ at any radius.
 
 \item The local oxygen abundance increases with color index $g-r$ of the galaxy
 outside the inner part.
 
 \item The local oxygen abundance depends on the central surface brightness of the disk $\Sigma_0$ 
 only in the outer region of the galaxy. 
The correlation between local oxygen abundance and surface brightness in the outer region 
of galaxies, in addition to the flattening of the radial metallicity gradient in 
the massive galaxies, can be considered as confirmation of the inside-out galaxy formation 
scenario \citep{Kepner1999,Pilkington2012}.

\end{enumerate}

\section*{Acknowledgements}

We are grateful to the referee for their constructive comments. 
I.A.Z. thanks the German Academic Exchange Service (DAAD) for financial support.
I.A.Z. and L.S.P.\
acknowledge support within the framework of Sonderforschungsbereich
(SFB 881) on ``The Milky Way System'' (especially subproject A5),
which is funded by the German Research Foundation (DFG).  L.S.P.\ and
I.A.Z.\ thank the Astronomisches Rechen-Institut at the
Universit\"{a}t Heidelberg,  where this investigation was carried out,
for the hospitality. \\
I.A.Z. acknowledges the support by the Ukrainian National Grid 
program (project 400Kt) of the NAS of Ukraine. \\
This study uses data provided by the Calar Alto Legacy Integral Field 
Area (CALIFA) survey (http://califa.caha.es/).
Based on observations collected at the Centro Astronomico Hispano Aleman 
(CAHA) at Calar Alto, operated jointly by the Max-Planck-Institut fur 
Astronomie and the Instituto de Astrofisica de Andalucia (CSIC).

\bibliography{AA.bib}

\begin{thebibliography}{91}
\expandafter\ifx\csname natexlab\endcsname\relax\def\natexlab#1{#1}\fi

\bibitem[{{Abraham} \& {Merrifield}(2000)}]{Abraham2000}
{Abraham}, R.~G. \& {Merrifield}, M.~R. 2000, \aj, 120, 2835

\bibitem[{{Aguerri}(1999)}]{Aguerri1999}
{Aguerri}, J.~A.~L. 1999, \aap, 351, 43

\bibitem[{{Aguerri} {et~al.}(1998){Aguerri}, {Beckman}, \&
  {Prieto}}]{Aguerri1998}
{Aguerri}, J.~A.~L., {Beckman}, J.~E., \& {Prieto}, M. 1998, \aj, 116, 2136

\bibitem[{{Aguerri} {et~al.}(2000){Aguerri}, {Mu{\~n}oz-Tu{\~n}{\'o}n},
  {Varela}, \& {Prieto}}]{Aguerri2000}
{Aguerri}, J.~A.~L., {Mu{\~n}oz-Tu{\~n}{\'o}n}, C., {Varela}, A.~M., \&
  {Prieto}, M. 2000, \aap, 361, 841

\bibitem[{{Asari} {et~al.}(2007){Asari}, {Cid Fernandes}, {Stasi{\'n}ska},
  {Torres-Papaqui}, {Mateus}, {Sodr{\'e}}, {Schoenell}, \& {Gomes}}]{Asari2007}
{Asari}, N.~V., {Cid Fernandes}, R., {Stasi{\'n}ska}, G., {et~al.} 2007,
  \mnras, 381, 263

\bibitem[{{Athanassoula}(1992)}]{Athanassoula1992}
{Athanassoula}, E. 1992, \mnras, 259, 345

\bibitem[{{Baldwin} {et~al.}(1981){Baldwin}, {Phillips}, \& {Terlevich}}]{BPT}
{Baldwin}, J.~A., {Phillips}, M.~M., \& {Terlevich}, R. 1981, \pasp, 93, 5

\bibitem[{{Belfiore} {et~al.}(2015){Belfiore}, {Maiolino}, {Bundy}, {Thomas},
  {Maraston}, {Wilkinson}, {S{\'a}nchez}, {Bershady}, {Blanc}, {Bothwell},
  {Cales}, {Coccato}, {Drory}, {Emsellem}, {Fu}, {Gelfand}, {Law}, {Masters},
  {Parejko}, {Tremonti}, {Wake}, {Weijmans}, {Yan}, {Xiao}, {Zhang}, {Zheng},
  {Bizyaev}, {Kinemuchi}, {Oravetz}, \& {Simmons}}]{Belfiore2015}
{Belfiore}, F., {Maiolino}, R., {Bundy}, K., {et~al.} 2015, \mnras, 449, 867

\bibitem[{{Belfiore} {et~al.}(2017){Belfiore}, {Maiolino}, {Tremonti},
  {S{\'a}nchez}, {Bundy}, {Bershady}, {Westfall}, {Lin}, {Drory}, {Boquien},
  {Thomas}, \& {Brinkmann}}]{Belfiore2017}
{Belfiore}, F., {Maiolino}, R., {Tremonti}, C., {et~al.} 2017, \mnras, 469, 151

\bibitem[{{Belley} \& {Roy}(1992)}]{Belley1992}
{Belley}, J. \& {Roy}, J.-R. 1992, \apjs, 78, 61

\bibitem[{{Blanton} {et~al.}(2017){Blanton}, {Bershady}, {Abolfathi},
  {Albareti}, {Allende Prieto}, {Almeida}, {Alonso-Garc{\'{\i}}a}, {Anders},
  {Anderson}, {Andrews}, \& et~al.}]{Blanton2017}
{Blanton}, M.~R., {Bershady}, M.~A., {Abolfathi}, B., {et~al.} 2017, \aj, 154,
  28

\bibitem[{{Blanton} \& {Roweis}(2007)}]{Blanton2007}
{Blanton}, M.~R. \& {Roweis}, S. 2007, \aj, 133, 734

\bibitem[{{Bresolin} {et~al.}(2012){Bresolin}, {Kennicutt}, \&
  {Ryan-Weber}}]{Bresolin2012}
{Bresolin}, F., {Kennicutt}, R.~C., \& {Ryan-Weber}, E. 2012, \apj, 750, 122

\bibitem[{{Bresolin} {et~al.}(2009){Bresolin}, {Ryan-Weber}, {Kennicutt}, \&
  {Goddard}}]{Bresolin2009}
{Bresolin}, F., {Ryan-Weber}, E., {Kennicutt}, R.~C., \& {Goddard}, Q. 2009,
  \apj, 695, 580

\bibitem[{{Bruzual} \& {Charlot}(2003)}]{BC03}
{Bruzual}, G. \& {Charlot}, S. 2003, \mnras, 344, 1000

\bibitem[{{Bundy} {et~al.}(2015){Bundy}, {Bershady}, {Law}, {Yan}, {Drory},
  {MacDonald}, {Wake}, {Cherinka}, {S{\'a}nchez-Gallego}, {Weijmans}, {Thomas},
  {Tremonti}, {Masters}, {Coccato}, {Diamond-Stanic}, {Arag{\'o}n-Salamanca},
  {Avila-Reese}, {Badenes}, {Falc{\'o}n-Barroso}, {Belfiore}, {Bizyaev},
  {Blanc}, {Bland-Hawthorn}, {Blanton}, {Brownstein}, {Byler}, {Cappellari},
  {Conroy}, {Dutton}, {Emsellem}, {Etherington}, {Frinchaboy}, {Fu}, {Gunn},
  {Harding}, {Johnston}, {Kauffmann}, {Kinemuchi}, {Klaene}, {Knapen},
  {Leauthaud}, {Li}, {Lin}, {Maiolino}, {Malanushenko}, {Malanushenko}, {Mao},
  {Maraston}, {McDermid}, {Merrifield}, {Nichol}, {Oravetz}, {Pan}, {Parejko},
  {Sanchez}, {Schlegel}, {Simmons}, {Steele}, {Steinmetz}, {Thanjavur},
  {Thompson}, {Tinker}, {van den Bosch}, {Westfall}, {Wilkinson}, {Wright},
  {Xiao}, \& {Zhang}}]{Bundy2015}
{Bundy}, K., {Bershady}, M.~A., {Law}, D.~R., {et~al.} 2015, \apj, 798, 7

\bibitem[{{Buta} {et~al.}(2003){Buta}, {Block}, \& {Knapen}}]{Buta2003}
{Buta}, R., {Block}, D.~L., \& {Knapen}, J.~H. 2003, \aj, 126, 1148

\bibitem[{{Cardelli} {et~al.}(1989){Cardelli}, {Clayton}, \& {Mathis}}]{CCM}
{Cardelli}, J.~A., {Clayton}, G.~C., \& {Mathis}, J.~S. 1989, \apj, 345, 245

\bibitem[{{Chapelon} {et~al.}(1999){Chapelon}, {Contini}, \&
  {Davoust}}]{Chapelon1999}
{Chapelon}, S., {Contini}, T., \& {Davoust}, E. 1999, \aap, 345, 81

\bibitem[{{Cheung} {et~al.}(2015){Cheung}, {Conroy}, {Athanassoula}, {Bell},
  {Bosma}, {Cardamone}, {Faber}, {Koo}, {Lintott}, {Masters}, {Melvin},
  {Simmons}, \& {Willett}}]{Cheung2015}
{Cheung}, E., {Conroy}, C., {Athanassoula}, E., {et~al.} 2015, \apj, 807, 36

\bibitem[{{Cid Fernandes} {et~al.}(2005){Cid Fernandes}, {Mateus}, {Sodr{\'e}},
  {Stasi{\'n}ska}, \& {Gomes}}]{CidFernandes2005}
{Cid Fernandes}, R., {Mateus}, A., {Sodr{\'e}}, L., {Stasi{\'n}ska}, G., \&
  {Gomes}, J.~M. 2005, \mnras, 358, 363

\bibitem[{{Combes} \& {Sanders}(1981)}]{Combes1981}
{Combes}, F. \& {Sanders}, R.~H. 1981, \aap, 96, 164

\bibitem[{{Di Matteo} {et~al.}(2013){Di Matteo}, {Haywood}, {Combes},
  {Semelin}, \& {Snaith}}]{DiMatteo2013}
{Di Matteo}, P., {Haywood}, M., {Combes}, F., {Semelin}, B., \& {Snaith}, O.~N.
  2013, \aap, 553, A102

\bibitem[{{D{\'{\i}}az-Garc{\'{\i}}a}
  {et~al.}(2016){D{\'{\i}}az-Garc{\'{\i}}a}, {Salo}, {Laurikainen}, \&
  {Herrera-Endoqui}}]{DiazGarcia2016}
{D{\'{\i}}az-Garc{\'{\i}}a}, S., {Salo}, H., {Laurikainen}, E., \&
  {Herrera-Endoqui}, M. 2016, \aap, 587, A160

\bibitem[{{Erwin}(2005)}]{Erwin2005}
{Erwin}, P. 2005, \mnras, 364, 283

\bibitem[{{Garcia-G{\'o}mez} {et~al.}(2017){Garcia-G{\'o}mez}, {Athanassoula},
  {Barber{\`a}}, \& {Bosma}}]{GarciaGomez2017}
{Garcia-G{\'o}mez}, C., {Athanassoula}, E., {Barber{\`a}}, C., \& {Bosma}, A.
  2017, \aap, 601, A132

\bibitem[{{Garnett}(1998)}]{Garnett1998}
{Garnett}, D.~R. 1998, in Revista Mexicana de Astronomia y Astrofisica
  Conference Series, Vol.~7, 58

\bibitem[{{Goddard} {et~al.}(2011){Goddard}, {Bresolin}, {Kennicutt},
  {Ryan-Weber}, \& {Rosales-Ortega}}]{Goddard2011}
{Goddard}, Q.~E., {Bresolin}, F., {Kennicutt}, R.~C., {Ryan-Weber}, E.~V., \&
  {Rosales-Ortega}, F.~F. 2011, \mnras, 412, 1246

\bibitem[{{Grand} \& {Kawata}(2016)}]{Grand2016}
{Grand}, R.~J.~J. \& {Kawata}, D. 2016, Astronomische Nachrichten, 337, 957

\bibitem[{{Haffner} {et~al.}(2009){Haffner}, {Dettmar}, {Beckman}, {Wood},
  {Slavin}, {Giammanco}, {Madsen}, {Zurita}, \& {Reynolds}}]{Haffner2009}
{Haffner}, L.~M., {Dettmar}, R.-J., {Beckman}, J.~E., {et~al.} 2009, Reviews of
  Modern Physics, 81, 969

\bibitem[{{Ho} {et~al.}(2015){Ho}, {Kudritzki}, {Kewley}, {Zahid}, {Dopita},
  {Bresolin}, \& {Rupke}}]{Ho2015}
{Ho}, I.-T., {Kudritzki}, R.-P., {Kewley}, L.~J., {et~al.} 2015, \mnras, 448,
  2030

\bibitem[{{Izotov} {et~al.}(1994){Izotov}, {Thuan}, \&
  {Lipovetsky}}]{Izotov1994}
{Izotov}, Y.~I., {Thuan}, T.~X., \& {Lipovetsky}, V.~A. 1994, \apj, 435, 647

\bibitem[{{Kauffmann} {et~al.}(2003){Kauffmann}, {Heckman}, {Tremonti},
  {Brinchmann}, {Charlot}, {White}, {Ridgway}, {Brinkmann}, {Fukugita}, {Hall},
  {Ivezi{\'c}}, {Richards}, \& {Schneider}}]{Kauffmann2003}
{Kauffmann}, G., {Heckman}, T.~M., {Tremonti}, C., {et~al.} 2003, \mnras, 346,
  1055

\bibitem[{{Kepner}(1999)}]{Kepner1999}
{Kepner}, J.~V. 1999, \apj, 520, 59

\bibitem[{{Kewley} \& {Ellison}(2008)}]{Kewley2008}
{Kewley}, L.~J. \& {Ellison}, S.~L. 2008, \apj, 681, 1183

\bibitem[{{Lacerda} {et~al.}(2018){Lacerda}, {Cid Fernandes}, {Couto},
  {Stasi{\'n}ska}, {Garc{\'{\i}}a-Benito}, {Vale Asari}, {P{\'e}rez},
  {Gonz{\'a}lez Delgado}, {S{\'a}nchez}, \& {de Amorim}}]{Lacerda2018}
{Lacerda}, E.~A.~D., {Cid Fernandes}, R., {Couto}, G.~S., {et~al.} 2018,
  \mnras, 474, 3727

\bibitem[{{Laurikainen} \& {Salo}(2002)}]{Laurikainen2002}
{Laurikainen}, E. \& {Salo}, H. 2002, \mnras, 337, 1118

\bibitem[{{Lian} {et~al.}(2018){Lian}, {Thomas}, {Maraston}, {Goddard},
  {Parikh}, {Fern{\'a}ndez-Trincado}, {Roman-Lopes}, {Rong}, {Tang}, \&
  {Yan}}]{Lian2018}
{Lian}, J., {Thomas}, D., {Maraston}, C., {et~al.} 2018, \mnras, 476, 3883

\bibitem[{{Marinova} \& {Jogee}(2007)}]{Marinova2007}
{Marinova}, I. \& {Jogee}, S. 2007, \apj, 659, 1176

\bibitem[{{Martin}(1995)}]{Martin1995}
{Martin}, P. 1995, \aj, 109, 2428

\bibitem[{{Martin} \& {Roy}(1994)}]{Martin1994}
{Martin}, P. \& {Roy}, J.-R. 1994, \apj, 424, 599

\bibitem[{{Martinet} \& {Friedli}(1997)}]{Martinet1997}
{Martinet}, L. \& {Friedli}, D. 1997, \aap, 323, 363

\bibitem[{{Mateus} {et~al.}(2006){Mateus}, {Sodr{\'e}}, {Cid Fernandes},
  {Stasi{\'n}ska}, {Schoenell}, \& {Gomes}}]{Mateus2006}
{Mateus}, A., {Sodr{\'e}}, L., {Cid Fernandes}, R., {et~al.} 2006, \mnras, 370,
  721

\bibitem[{{Mayor}(1976)}]{Mayor1976}
{Mayor}, M. 1976, in IAU Symposium, Vol.~72, Abundance Effects in
  Classification, ed. B.~{Hauck}, P.~C. {Keenan}, \& W.~W. {Morgan}, 207

\bibitem[{{McCall} {et~al.}(1985){McCall}, {Rybski}, \& {Shields}}]{McCall1985}
{McCall}, M.~L., {Rybski}, P.~M., \& {Shields}, G.~A. 1985, \apjs, 57, 1

\bibitem[{{Minchev} {et~al.}(2013){Minchev}, {Chiappini}, \&
  {Martig}}]{Minchev2013}
{Minchev}, I., {Chiappini}, C., \& {Martig}, M. 2013, \aap, 558, A9

\bibitem[{{Minchev} {et~al.}(2014){Minchev}, {Chiappini}, \&
  {Martig}}]{Minchev2014}
{Minchev}, I., {Chiappini}, C., \& {Martig}, M. 2014, \aap, 572, A92

\bibitem[{{Minchev} \& {Famaey}(2010)}]{Minchev2010}
{Minchev}, I. \& {Famaey}, B. 2010, \apj, 722, 112

\bibitem[{{Patterson} {et~al.}(2012){Patterson}, {Walterbos}, {Kennicutt},
  {Chiappini}, \& {Thilker}}]{Patterson2012}
{Patterson}, M.~T., {Walterbos}, R.~A.~M., {Kennicutt}, R.~C., {Chiappini}, C.,
  \& {Thilker}, D.~A. 2012, \mnras, 422, 401

\bibitem[{{Peimbert}(1979)}]{Peimbert1979}
{Peimbert}, M. 1979, in IAU Symposium, Vol.~84, The Large-Scale Characteristics
  of the Galaxy, ed. W.~B. {Burton}, 307--315

\bibitem[{{Peng} {et~al.}(2002){Peng}, {Ho}, {Impey}, \& {Rix}}]{Peng2002}
{Peng}, C.~Y., {Ho}, L.~C., {Impey}, C.~D., \& {Rix}, H.-W. 2002, \aj, 124, 266

\bibitem[{{Peng} {et~al.}(2010){Peng}, {Ho}, {Impey}, \& {Rix}}]{Peng2010}
{Peng}, C.~Y., {Ho}, L.~C., {Impey}, C.~D., \& {Rix}, H.-W. 2010, \aj, 139,
  2097

\bibitem[{{Pilkington} {et~al.}(2012){Pilkington}, {Few}, {Gibson}, {Calura},
  {Michel-Dansac}, {Thacker}, {Moll{\'a}}, {Matteucci}, {Rahimi}, {Kawata},
  {Kobayashi}, {Brook}, {Stinson}, {Couchman}, {Bailin}, \&
  {Wadsley}}]{Pilkington2012}
{Pilkington}, K., {Few}, C.~G., {Gibson}, B.~K., {et~al.} 2012, \aap, 540, A56

\bibitem[{{Pilyugin} \& {Grebel}(2016)}]{PilyuginGrebel2016}
{Pilyugin}, L.~S. \& {Grebel}, E.~K. 2016, \mnras, 457, 3678

\bibitem[{{Pilyugin} {et~al.}(2014{\natexlab{a}}){Pilyugin}, {Grebel}, \&
  {Kniazev}}]{Pilyugin2014a}
{Pilyugin}, L.~S., {Grebel}, E.~K., \& {Kniazev}, A.~Y. 2014{\natexlab{a}},
  \aj, 147, 131

\bibitem[{{Pilyugin} {et~al.}(2014{\natexlab{b}}){Pilyugin}, {Grebel},
  {Zinchenko}, \& {Kniazev}}]{Pilyugin2014b}
{Pilyugin}, L.~S., {Grebel}, E.~K., {Zinchenko}, I.~A., \& {Kniazev}, A.~Y.
  2014{\natexlab{b}}, \aj, 148, 134

\bibitem[{{Pilyugin} {et~al.}(2018){Pilyugin}, {Grebel}, {Zinchenko},
  {Nefedyev}, {Shulga}, {Wei}, \& {Berczik}}]{Pilyugin2018}
{Pilyugin}, L.~S., {Grebel}, E.~K., {Zinchenko}, I.~A., {et~al.} 2018, \aap,
  613, A1

\bibitem[{{Pilyugin} {et~al.}(2017){Pilyugin}, {Grebel}, {Zinchenko},
  {Nefedyev}, \& {V{\'{\i}}lchez}}]{Pilyugin2017}
{Pilyugin}, L.~S., {Grebel}, E.~K., {Zinchenko}, I.~A., {Nefedyev}, Y.~A., \&
  {V{\'{\i}}lchez}, J.~M. 2017, \aap, 608, A127

\bibitem[{{Prantzos} \& {Boissier}(2000)}]{Prantzos2000}
{Prantzos}, N. \& {Boissier}, S. 2000, \mnras, 313, 338

\bibitem[{{Rich} {et~al.}(2012){Rich}, {Torrey}, {Kewley}, {Dopita}, \&
  {Rupke}}]{Rich2012}
{Rich}, J.~A., {Torrey}, P., {Kewley}, L.~J., {Dopita}, M.~A., \& {Rupke},
  D.~S.~N. 2012, \apj, 753, 5

\bibitem[{{Rosales-Ortega} {et~al.}(2011){Rosales-Ortega}, {D{\'{\i}}az},
  {Kennicutt}, \& {S{\'a}nchez}}]{RosalesOrtega2011}
{Rosales-Ortega}, F.~F., {D{\'{\i}}az}, A.~I., {Kennicutt}, R.~C., \&
  {S{\'a}nchez}, S.~F. 2011, \mnras, 415, 2439

\bibitem[{{Roy}(1996)}]{Roy1996}
{Roy}, J.-R. 1996, in Astronomical Society of the Pacific Conference Series,
  Vol.~91, IAU Colloq. 157: Barred Galaxies, ed. R.~{Buta}, D.~A. {Crocker}, \&
  B.~G. {Elmegreen}, 63

\bibitem[{{Ruiz-Lara} {et~al.}(2017){Ruiz-Lara}, {P{\'e}rez}, {Florido},
  {S{\'a}nchez-Bl{\'a}zquez}, {M{\'e}ndez-Abreu}, {S{\'a}nchez-Menguiano},
  {S{\'a}nchez}, {Lyubenova}, {Falc{\'o}n-Barroso}, {van de Ven}, {Marino}, {de
  Lorenzo-C{\'a}ceres}, {Catal{\'a}n-Torrecilla}, {Costantin},
  {Bland-Hawthorn}, {Galbany}, {Garc{\'{\i}}a-Benito}, {Husemann}, {Kehrig},
  {M{\'a}rquez}, {Mast}, {Walcher}, {Zibetti}, {Ziegler}, \& {Califa
  Team}}]{RuizLara2017}
{Ruiz-Lara}, T., {P{\'e}rez}, I., {Florido}, E., {et~al.} 2017, \aap, 604, A4

\bibitem[{{Ryder}(1995)}]{Ryder1995}
{Ryder}, S.~D. 1995, \apj, 444, 610

\bibitem[{{Saha} \& {Naab}(2013)}]{Saha2013}
{Saha}, K. \& {Naab}, T. 2013, \mnras, 434, 1287

\bibitem[{{Sakhibov} {et~al.}(2018){Sakhibov}, {Zinchenko}, {Pilyugin},
  {Grebel}, {Just}, \& {V{\'{\i}}lchez}}]{Sakhibov2018}
{Sakhibov}, F., {Zinchenko}, I.~A., {Pilyugin}, L.~S., {et~al.} 2018, \mnras,
  474, 1657

\bibitem[{{S{\'a}nchez} {et~al.}(2016){S{\'a}nchez}, {Garc{\'{\i}}a-Benito},
  {Zibetti}, {Walcher}, {Husemann}, {Mendoza}, {Galbany}, {Falc{\'o}n-Barroso},
  {Mast}, {Aceituno}, {Aguerri}, {Alves}, {Amorim}, {Ascasibar},
  {Barrado-Navascues}, {Barrera-Ballesteros}, {Bekerait{\`e}},
  {Bland-Hawthorn}, {Cano D{\'{\i}}az}, {Cid Fernandes}, {Cavichia}, {Cortijo},
  {Dannerbauer}, {Demleitner}, {D{\'{\i}}az}, {Dettmar}, {de
  Lorenzo-C{\'a}ceres}, {del Olmo}, {Galazzi}, {Garc{\'{\i}}a-Lorenzo}, {Gil de
  Paz}, {Gonz{\'a}lez Delgado}, {Holmes}, {Igl{\'e}sias-P{\'a}ramo}, {Kehrig},
  {Kelz}, {Kennicutt}, {Kleemann}, {Lacerda}, {L{\'o}pez Fern{\'a}ndez},
  {L{\'o}pez S{\'a}nchez}, {Lyubenova}, {Marino}, {M{\'a}rquez},
  {Mendez-Abreu}, {Moll{\'a}}, {Monreal-Ibero}, {Ortega Minakata},
  {Torres-Papaqui}, {P{\'e}rez}, {Rosales-Ortega}, {Roth},
  {S{\'a}nchez-Bl{\'a}zquez}, {Schilling}, {Spekkens}, {Vale Asari}, {van den
  Bosch}, {van de Ven}, {Vilchez}, {Wild}, {Wisotzki}, {Y{\i}ld{\i}r{\i}m}, \&
  {Ziegler}}]{Sanchez2016}
{S{\'a}nchez}, S.~F., {Garc{\'{\i}}a-Benito}, R., {Zibetti}, S., {et~al.} 2016,
  \aap, 594, A36

\bibitem[{{S{\'a}nchez} {et~al.}(2012){S{\'a}nchez}, {Kennicutt}, {Gil de Paz},
  {van de Ven}, {V{\'{\i}}lchez}, {Wisotzki}, {Walcher}, {Mast}, {Aguerri},
  {Albiol-P{\'e}rez}, {Alonso-Herrero}, {Alves}, {Bakos}, {Bart{\'a}kov{\'a}},
  {Bland-Hawthorn}, {Boselli}, {Bomans}, {Castillo-Morales}, {Cortijo-Ferrero},
  {de Lorenzo-C{\'a}ceres}, {Del Olmo}, {Dettmar}, {D{\'{\i}}az}, {Ellis},
  {Falc{\'o}n-Barroso}, {Flores}, {Gallazzi}, {Garc{\'{\i}}a-Lorenzo},
  {Gonz{\'a}lez Delgado}, {Gruel}, {Haines}, {Hao}, {Husemann},
  {Igl{\'e}sias-P{\'a}ramo}, {Jahnke}, {Johnson}, {Jungwiert}, {Kalinova},
  {Kehrig}, {Kupko}, {L{\'o}pez-S{\'a}nchez}, {Lyubenova}, {Marino},
  {M{\'a}rmol-Queralt{\'o}}, {M{\'a}rquez}, {Masegosa}, {Meidt},
  {Mendez-Abreu}, {Monreal-Ibero}, {Montijo}, {Mour{\~a}o}, {Palacios-Navarro},
  {Papaderos}, {Pasquali}, {Peletier}, {P{\'e}rez}, {P{\'e}rez}, {Quirrenbach},
  {Rela{\~n}o}, {Rosales-Ortega}, {Roth}, {Ruiz-Lara},
  {S{\'a}nchez-Bl{\'a}zquez}, {Sengupta}, {Singh}, {Stanishev}, {Trager},
  {Vazdekis}, {Viironen}, {Wild}, {Zibetti}, \& {Ziegler}}]{Sanchez2012}
{S{\'a}nchez}, S.~F., {Kennicutt}, R.~C., {Gil de Paz}, A., {et~al.} 2012,
  \aap, 538, A8

\bibitem[{{S{\'a}nchez} {et~al.}(2014{\natexlab{a}}){S{\'a}nchez},
  {Rosales-Ortega}, {Iglesias-P{\'a}ramo}, {Moll{\'a}}, {Barrera-Ballesteros},
  {Marino}, {P{\'e}rez}, {S{\'a}nchez-Blazquez}, {Gonz{\'a}lez Delgado}, {Cid
  Fernandes}, {de Lorenzo-C{\'a}ceres}, {Mendez-Abreu}, {Galbany},
  {Falcon-Barroso}, {Miralles-Caballero}, {Husemann}, {Garc{\'{\i}}a-Benito},
  {Mast}, {Walcher}, {Gil de Paz}, {Garc{\'{\i}}a-Lorenzo}, {Jungwiert},
  {V{\'{\i}}lchez}, {J{\'{\i}}lkov{\'a}}, {Lyubenova}, {Cortijo-Ferrero},
  {D{\'{\i}}az}, {Wisotzki}, {M{\'a}rquez}, {Bland-Hawthorn}, {Ellis}, {van de
  Ven}, {Jahnke}, {Papaderos}, {Gomes}, {Mendoza}, \&
  {L{\'o}pez-S{\'a}nchez}}]{Sanchez2013}
{S{\'a}nchez}, S.~F., {Rosales-Ortega}, F.~F., {Iglesias-P{\'a}ramo}, J.,
  {et~al.} 2014{\natexlab{a}}, \aap, 563, A49

\bibitem[{{S{\'a}nchez} {et~al.}(2014{\natexlab{b}}){S{\'a}nchez},
  {Rosales-Ortega}, {Iglesias-P{\'a}ramo}, {Moll{\'a}}, {Barrera-Ballesteros},
  {Marino}, {P{\'e}rez}, {S{\'a}nchez-Blazquez}, {Gonz{\'a}lez Delgado}, {Cid
  Fernandes}, {de Lorenzo-C{\'a}ceres}, {Mendez-Abreu}, {Galbany},
  {Falcon-Barroso}, {Miralles-Caballero}, {Husemann}, {Garc{\'{\i}}a-Benito},
  {Mast}, {Walcher}, {Gil de Paz}, {Garc{\'{\i}}a-Lorenzo}, {Jungwiert},
  {V{\'{\i}}lchez}, {J{\'{\i}}lkov{\'a}}, {Lyubenova}, {Cortijo-Ferrero},
  {D{\'{\i}}az}, {Wisotzki}, {M{\'a}rquez}, {Bland-Hawthorn}, {Ellis}, {van de
  Ven}, {Jahnke}, {Papaderos}, {Gomes}, {Mendoza}, \&
  {L{\'o}pez-S{\'a}nchez}}]{Sanchez2014}
{S{\'a}nchez}, S.~F., {Rosales-Ortega}, F.~F., {Iglesias-P{\'a}ramo}, J.,
  {et~al.} 2014{\natexlab{b}}, \aap, 563, A49

\bibitem[{{S{\'a}nchez-Bl{\'a}zquez} {et~al.}(2014){S{\'a}nchez-Bl{\'a}zquez},
  {Rosales-Ortega}, {M{\'e}ndez-Abreu}, {P{\'e}rez}, {S{\'a}nchez}, {Zibetti},
  {Aguerri}, {Bland-Hawthorn}, {Catal{\'a}n-Torrecilla}, {Cid Fernandes}, {de
  Amorim}, {de Lorenzo-Caceres}, {Falc{\'o}n-Barroso}, {Galazzi},
  {Garc{\'{\i}}a Benito}, {Gil de Paz}, {Gonz{\'a}lez Delgado}, {Husemann},
  {Iglesias-P{\'a}ramo}, {Jungwiert}, {Marino}, {M{\'a}rquez}, {Mast},
  {Mendoza}, {Moll{\'a}}, {Papaderos}, {Ruiz-Lara}, {van de Ven}, {Walcher}, \&
  {Wisotzki}}]{Sanchez-Blazquez2014}
{S{\'a}nchez-Bl{\'a}zquez}, P., {Rosales-Ortega}, F.~F., {M{\'e}ndez-Abreu},
  J., {et~al.} 2014, \aap, 570, A6

\bibitem[{{S{\'a}nchez-Menguiano} {et~al.}(2017){S{\'a}nchez-Menguiano},
  {S{\'a}nchez}, {P{\'e}rez}, {Debattista}, {Ruiz-Lara}, {Florido}, {Cavichia},
  {Galbany}, {Marino}, {Mast}, {S{\'a}nchez-Bl{\'a}zquez}, {M{\'e}ndez-Abreu},
  {de Lorenzo-C{\'a}ceres}, {Catal{\'a}n-Torrecilla}, {Cano-D{\'{\i}}az},
  {M{\'a}rquez}, {McIntosh}, {Ascasibar}, {Garc{\'{\i}}a-Benito}, {G{\'o}nzalez
  Delgado}, {Kehrig}, {L{\'o}pez-S{\'a}nchez}, {Moll{\'a}}, {Bland-Hawthorn},
  {Walcher}, \& {Costantin}}]{Sanchez-Menguiano2017}
{S{\'a}nchez-Menguiano}, L., {S{\'a}nchez}, S.~F., {P{\'e}rez}, I., {et~al.}
  2017, \aap, 603, A113

\bibitem[{{S{\'a}nchez-Menguiano} {et~al.}(2016){S{\'a}nchez-Menguiano},
  {S{\'a}nchez}, {P{\'e}rez}, {Garc{\'{\i}}a-Benito}, {Husemann}, {Mast},
  {Mendoza}, {Ruiz-Lara}, {Ascasibar}, {Bland-Hawthorn}, {Cavichia},
  {D{\'{\i}}az}, {Florido}, {Galbany}, {G{\'o}nzalez Delgado}, {Kehrig},
  {Marino}, {M{\'a}rquez}, {Masegosa}, {M{\'e}ndez-Abreu}, {Moll{\'a}}, {Del
  Olmo}, {P{\'e}rez}, {S{\'a}nchez-Bl{\'a}zquez}, {Stanishev}, {Walcher},
  {L{\'o}pez-S{\'a}nchez}, \& {Califa Collaboration}}]{SanchezMenguiano2016}
{S{\'a}nchez-Menguiano}, L., {S{\'a}nchez}, S.~F., {P{\'e}rez}, I., {et~al.}
  2016, \aap, 587, A70

\bibitem[{{S{\'a}nchez-Menguiano} {et~al.}(2018){S{\'a}nchez-Menguiano},
  {S{\'a}nchez}, {P{\'e}rez}, {Ruiz-Lara}, {Galbany}, {Anderson},
  {Kr{\"u}hler}, {Kuncarayakti}, \& {Lyman}}]{SanchezMenguiano2018}
{S{\'a}nchez-Menguiano}, L., {S{\'a}nchez}, S.~F., {P{\'e}rez}, I., {et~al.}
  2018, \aap, 609, A119

\bibitem[{{Schlafly} \& {Finkbeiner}(2011)}]{Schlafly2011}
{Schlafly}, E.~F. \& {Finkbeiner}, D.~P. 2011, \apj, 737, 103

\bibitem[{{Schlegel} {et~al.}(1998){Schlegel}, {Finkbeiner}, \&
  {Davis}}]{Schlegel1998}
{Schlegel}, D.~J., {Finkbeiner}, D.~P., \& {Davis}, M. 1998, \apj, 500, 525

\bibitem[{{Sch{\"o}nrich} \& {Binney}(2009)}]{Schonrich2009}
{Sch{\"o}nrich}, R. \& {Binney}, J. 2009, \mnras, 396, 203

\bibitem[{{Searle}(1971)}]{Searle1971}
{Searle}, L. 1971, \apj, 168, 327

\bibitem[{{Sellwood} \& {Binney}(2002)}]{Sellwood2002}
{Sellwood}, J.~A. \& {Binney}, J.~J. 2002, \mnras, 336, 785

\bibitem[{{Shields} \& {Searle}(1978)}]{Shields1978}
{Shields}, G.~A. \& {Searle}, L. 1978, \apj, 222, 821

\bibitem[{{Storey} \& {Zeippen}(2000)}]{Storey2000}
{Storey}, P.~J. \& {Zeippen}, C.~J. 2000, \mnras, 312, 813

\bibitem[{{Thuan} {et~al.}(2010){Thuan}, {Pilyugin}, \&
  {Zinchenko}}]{Thuan2010}
{Thuan}, T.~X., {Pilyugin}, L.~S., \& {Zinchenko}, I.~A. 2010, \apj, 712, 1029

\bibitem[{{Tissera} {et~al.}(2016){Tissera}, {Pedrosa}, {Sillero}, \&
  {Vilchez}}]{Tissera2016}
{Tissera}, P.~B., {Pedrosa}, S.~E., {Sillero}, E., \& {Vilchez}, J.~M. 2016,
  \mnras, 456, 2982

\bibitem[{{Tremonti} {et~al.}(2004){Tremonti}, {Heckman}, {Kauffmann},
  {Brinchmann}, {Charlot}, {White}, {Seibert}, {Peng}, {Schlegel}, {Uomoto},
  {Fukugita}, \& {Brinkmann}}]{Tremonti2004}
{Tremonti}, C.~A., {Heckman}, T.~M., {Kauffmann}, G., {et~al.} 2004, \apj, 613,
  898

\bibitem[{{Vila-Costas} \& {Edmunds}(1992)}]{Vila-Costas1992}
{Vila-Costas}, M.~B. \& {Edmunds}, M.~G. 1992, \mnras, 259, 121

\bibitem[{{Walcher} {et~al.}(2014){Walcher}, {Wisotzki}, {Bekerait{\'e}},
  {Husemann}, {Iglesias-P{\'a}ramo}, {Backsmann}, {Barrera Ballesteros},
  {Catal{\'a}n-Torrecilla}, {Cortijo}, {del Olmo}, {Garcia Lorenzo},
  {Falc{\'o}n-Barroso}, {Jilkova}, {Kalinova}, {Mast}, {Marino},
  {M{\'e}ndez-Abreu}, {Pasquali}, {S{\'a}nchez}, {Trager}, {Zibetti},
  {Aguerri}, {Alves}, {Bland-Hawthorn}, {Boselli}, {Castillo Morales}, {Cid
  Fernandes}, {Flores}, {Galbany}, {Gallazzi}, {Garc{\'{\i}}a-Benito}, {Gil de
  Paz}, {Gonz{\'a}lez-Delgado}, {Jahnke}, {Jungwiert}, {Kehrig}, {Lyubenova},
  {M{\'a}rquez Perez}, {Masegosa}, {Monreal Ibero}, {P{\'e}rez}, {Quirrenbach},
  {Rosales-Ortega}, {Roth}, {Sanchez-Blazquez}, {Spekkens}, {Tundo}, {van de
  Ven}, {Verheijen}, {Vilchez}, \& {Ziegler}}]{CALIFA2014}
{Walcher}, C.~J., {Wisotzki}, L., {Bekerait{\'e}}, S., {et~al.} 2014, \aap,
  569, A1

\bibitem[{{Werk} {et~al.}(2011){Werk}, {Putman}, {Meurer}, \&
  {Santiago-Figueroa}}]{Werk2011}
{Werk}, J.~K., {Putman}, M.~E., {Meurer}, G.~R., \& {Santiago-Figueroa}, N.
  2011, \apj, 735, 71

\bibitem[{{Wu} {et~al.}(2015){Wu}, {Kudritzki}, {Tully}, \& {Neill}}]{Wu2015}
{Wu}, P.-F., {Kudritzki}, R.-P., {Tully}, R.~B., \& {Neill}, J.~D. 2015, \apj,
  810, 151

\bibitem[{{Zaritsky} {et~al.}(1994){Zaritsky}, {Kennicutt}, \&
  {Huchra}}]{Zaritsky1994}
{Zaritsky}, D., {Kennicutt}, Jr., R.~C., \& {Huchra}, J.~P. 1994, \apj, 420, 87

\bibitem[{{Zinchenko} {et~al.}(2015){Zinchenko}, {Kniazev}, {Grebel}, \&
  {Pilyugin}}]{Zinchenko2015}
{Zinchenko}, I.~A., {Kniazev}, A.~Y., {Grebel}, E.~K., \& {Pilyugin}, L.~S.
  2015, \aap, 582, A35

\bibitem[{{Zinchenko} {et~al.}(2016){Zinchenko}, {Pilyugin}, {Grebel},
  {S{\'a}nchez}, \& {V{\'{\i}}lchez}}]{Zinchenko2016}
{Zinchenko}, I.~A., {Pilyugin}, L.~S., {Grebel}, E.~K., {S{\'a}nchez}, S.~F.,
  \& {V{\'{\i}}lchez}, J.~M. 2016, \mnras, 462, 2715

\end{thebibliography}

\end{document}